\documentclass[12pt]{iopart}
\usepackage{graphicx, epstopdf}
\usepackage{amsmath,amssymb,latexsym}
\usepackage{braket}

\usepackage{siunitx}
\DeclareSIUnit\gauss{G}
\DeclareSIUnit\torr{Torr}
\DeclareSIUnit\LabView{LV}
\DeclareSIUnit\pixel{px}
\DeclareSIUnit\inch{^{\prime\prime}}
\DeclareSIUnit\miliK{mK}

\newcommand{\tab}[1][1cm]{\hspace*{#1}}
 
\begin{document}

\title{Ramsey interferometry in three-level and five-level systems of $^{87}Rb$ Bose-Einstein condensates}

\author{Anushka Thenuwara$^{1,2}$, Andrei Sidorov$^{1}$}

\address{$^1$ Centre for Optical Sciences, Faculty of Science, Engineering and Technology, Swinburne University of Technology, John Street, Hawthorn, Victoria, Australia 3122
$^2$  School of Physics and Astronomy, Monash University Clayton Campus, Wellington Rd, Clayton, Victoria, Australia 3800}
\ead{anushka.thenuwara@monash.edu and asidorov@swin.edu.au}
\vspace{10pt}
\begin{indented}
\item[]July 2021
\end{indented}

\begin{abstract}
Our work here presents the analytical expressions for a typical Ramsey interferometric sequence for a three- and a five-level system. The analytical expressions are derived starting from the first principals of unitary time evolution operators. We focus on the three- and five-level systems because we propose a novel Ramsey interferometer created by a trapped two-state Bose-Einstein Condensate driven by dipole oscillations and gravitational sag. It involves the  $^{87}Rb$ atoms in states $\ket{F=2, m_F=+2}$ $(\ket{+2})$ and $\ket{F=2, m_F=+1}$ $(\ket{+1})$ of the $5 ^2S_{\frac{1}{2}}$ ground state. Though the interferometer focusses on the two-levels, the experimental readouts  involve all the five states in $F = 2$ hyperfine manifold. Therefore, the analytical derivation was first tested for three-levels and then expanded to five-levels. We developed the expressions for five-levels for greater analytical accuracy of the experimental scenario. This work provides a step-by-step outline for the derivation and methodology for the analytical expressions. These analytical formulae denote the population variation during Rabi and Ramsey oscillations for each state as well as the overall average for both the three- and five-level cases. The expressions are derived within the rotating wave approximation (RWA) under the equal Rabi condition. Further, by following the derivation methodology, these analytical expressions can be easily expanded for Ramsey sequences with unequal pulses, and Ramsey sequences with spin echo techniques.
\end{abstract}

%
%
%
%
%

\section{Introduction}
Following the work of I.I. Rabi, N.F. Ramsey \cite{Ramsey1990} significantly improved the Rabi method by using two oscillatory fields with a short pulse $\tau$ separated by a long free evolution time $T$ to study molecular resonances and demonstrated \num{0.6} times narrower linewidths. This work of N.F. Ramsey won him the Nobel prize in physics in \num{1989} and is named as Ramsey interferometry. This method provides the basis of the exquisite time standards of $Cs$ fountain clocks such as the NIM5 clock in China and the NIST-F1 in the USA that have uncertainties of \num{1.6e-15} \cite{Fang2015} and \num{0.97e-15} \cite{Heavner2005} respectively. Further, Ramsey interferometry allows sensitive measurements of local gravity \cite{Peters1999} and a Ramsey-type method with a spin-echo pulse (i.e. a $\pi$-pulse during the free evolution time $T$) allows precise measurements of the Newtonian gravitational constant $G = \SI{6.67191(99)e-11}{\m^3 \kg^{-1} \s^{-2}}$ \cite{Rosi2014}. This applied in multilevel systems allows to measure below the standard quantum limit \cite{Petrovic2013} where the phase difference between states are mapped onto the populations of states \cite{Anderson2009}.

The primary goal of this work is to develop analytical expressions to explore and analyse experimental data of a novel Ramsey interferometer created by a trapped two-state Bose-Einstein condensate (BEC) driven by dipole oscillations and gravitational sag. A BEC is formed in a pure compressed magnetic trap (CMT) via a cloud of $^{87}Rb$ atoms in state $\ket{F=2, m_F=+2}$ $(\ket{+2})$ of the $5 ^2S_{\frac{1}{2}}$ ground state, from which Rmasey interferometry is performed between states $\ket{+2}$ and $\ket{F=2, m_F=+1}$ $(\ket{+1})$. The state $\ket{+1}$ experiences a shallower radial trap with a larger gravitational sag; whereas, state $\ket{+2}$ experiences a tighter radial trap with a gravitational sag that is half of state $\ket{+1}$. Due to this, a superposition between the states $\ket{+1}$ and $\ket{+2}$ experiences multipath propagation resulting in an interference pattern. This may be utilised to measure local gravitational fields and measure inter-sate scattering lengths.

In previous works \cite{BiaynickaBirula1977,Cook1979,Anderson2010,Vitanov1997} analytical expressions for Rabi oscillations in multi-level system were attained based on the two-level atom (Equation \ref{eqn:Flvl}). In \cite{BiaynickaBirula1977} considers an N-level atom interacting with a near-resonant laser field in two fronts; where the energy separation between all N-states are equal (equal-Rabi), and the energy separation increases as a harmonic potential (harmonic-Rabi). It also shows the association with Chebyshev and Hermite polynomials. The work in \cite{Cook1979} incorporates the treatment of losses and \cite{Fujii2003, Fujii2004, Fujii2006} shows the importance of understanding Rabi oscillations of a multi-level system for quantum computation. These formulations are performed within the rotating wave approximation (RWA) which is valid when the coupling constant (Rabi frequency) is much smaller than the energy separation between the two levels \cite{Fujii2003}. In this work we derive analytical expressions for a three- and five-level systems for the full Ramsey sequence via the unitary time evolution operator formalism under the equal-Rabi and RWA conditions.

The equal-Rabi assumption is valid as the trap bottom of the harmonic oscillator potential of the experiment is about \SI{1}{\gauss} relating to about \SI{700}{\kHz} energy separation between the five states in the $F=2$ hyperfine manifold \cite{Daniel2019}. The Breit-Rabi formula \cite{Breit1931, Wu2014} indicates there is only a \SI{0.02}{\%} variation in energy between adjacent states spanning $\ket{+2}$ to $\ket{-2}$. Due to this the equal-Rabi model suffices and the harmonic-Rabi model only complicates the analysis. Further, the maximum experimental Rabi frequency is less than \SI{15}{\%} of that of the energy separation between states which justifies the RWA approximation. 

Within these considerations, Section \ref{three-level system} derives the analytical expressions for the unitary time evolution $\hat{U}$ for the three-level system via $\hat{U} = \sum_{i=1}^{n} e^{\frac{-i \lambda_i t}{\hbar}} \ket{V_i} \bra{V_i}$ (Equation \ref{eqn:UDerive}), where $\lambda_i$ are the eigenvalues and $\ket{V_i(t)}$ are the eigenvectors of the interaction Hamiltonian $\hat{H_I}$. Once the methodology is validated, analytical expressions for the five-level system are derived and presented in Section \ref{five-level system}. Mathematica was used to solve these analytically dense problems.

\section{Rabi and Ramsey analytical models for three-level system}\label{three-level system}
Consider a three-level system  with equal energy separation between adjacent states leading to $\omega_{\ket{+1}}-\omega_{\ket{0}}=\omega_{\ket{0}}-\omega_{\ket{-1}} = \omega_{Sep}$ being coupled to an external EM field with a frequency $\omega_{EM} = {\omega_{Sep}}-\Delta$ where $\Delta$ is the detuning. The Rabi frequency for this system follows the resonant Rabi frequency $\Omega_R$ in Equation \ref{eqn:Psi_Sys} where $\Omega_R = \frac{\bra{1}\hat{\mu}\ket{2}B_0}{\hbar}$. Here, the  magnetic dipole coupling between adjacent states are equal $\bra{+1}\hat{\mu}\ket{0}B_0 = \bra{0}\hat{\mu}\ket{-1}B_0$ due to the equal energy separation between states.

The equal-Rabi interaction Hamiltonian $\hat{H}_I$ for the three-level system in the rotating wave approximation (RWA) is shown below as adapted from \cite{Anderson2010, Vitanov1997}.
\begin{equation}
	\label{eqn:3LvlHI} 
	\hat{H}_I = \hbar 
	\left[
	\begin{array} {ccc} 
	\Delta & \frac{1}{\sqrt{2}}\Omega_R e^{-i\phi} & 0\\
	\frac{1}{\sqrt{2}}\Omega_R e^{i\phi} & 0 & \frac{1}{\sqrt{2}}\Omega_R e^{-i\phi}\\
	0 & \frac{1}{\sqrt{2}}\Omega_R e^{i\phi} & -\Delta\\ 
	\end{array}
	\right].
\end{equation}

Using standard means of solving the eigenvector/eigenvalue problem we find the eigenvalues $\lambda_i$ via $|\hat{H}_I-\lambda 1|_{det} = 0 $ and the eigenvectors $V_i$ via $(\hat{H}_I-\lambda 1)\vec{V}=0$ where $1$ is the identity matrix, leading to
\begin{equation}
	\label{eqn:3LvlEV} 
	\left[V_1\right]_{\lambda_1} =
	\left[
	\begin{array} {c}
	\frac{-\Omega_R}{\sqrt{2}\Omega_G}\\
	\frac{\Delta}{\Omega_G}\\
	\frac{\Omega_R}{\sqrt{2}\Omega_G}\\
	\end{array}
	\right]_{0},
	\left[V_2\right]_{\lambda_2} =
	\left[
	\begin{array} {c}
	\frac{1}{2}\left(-1+\frac{\Delta}{\Omega_G}\right)\\
	\frac{\Omega_R}{\sqrt{2}\Omega_G}\\
	\frac{1}{2}\left(-1-\frac{\Delta}{\Omega_G}\right)\\
	\end{array}
	\right]_{-\hbar \Omega_G},
	\left[V_3\right]_{\lambda_3} =
	\left[
	\begin{array} {c}
	\frac{1}{2}\left(1+\frac{\Delta}{\Omega_G}\right)\\
	\frac{\Omega_R}{\sqrt{2}\Omega_G}\\
	\frac{1}{2}\left(1-\frac{\Delta}{\Omega_G}\right)\\
	\end{array}
	\right]_{\hbar \Omega_G},
\end{equation}
where  $V_i$ is the eigenvector normalized to \num{1}, the eigenvalues are $\begin{bmatrix} \lambda_1 & \lambda_2 & \lambda_3 \end{bmatrix} ^T = \begin{bmatrix} 0 & -\hbar \Omega_G & \hbar \Omega_G \end{bmatrix} ^T$ and $\Omega_G = \sqrt{\Delta^2+\Omega_R^2}$ is the general Rabi frequency.

Once the eigenvalues and eigenvectors are obtained, the unitary time evolution operator $\hat{U}$ for the general case is $\hat{U} = \sum_{i=1}^{n} e^{\frac{-i \lambda_i t}{\hbar}} \ket{V_i} \bra{V_i}$. This expression is derived as $\hat{H}$ is independent of time in the RWA. However, via the unitary time evolution operator $\hat{U}$, we can use $\ket{\Psi(t)}= \hat{U} \ket{\Psi(0)}$ where $\hat{U} = e^{\frac{i\hat{H}t}{\hbar}}$ which facilitates an easier method to obtain analytical solutions. Therefore it is crucial to obtain an expression for $e^{\frac{i\hat{H}t}{\hbar}}$. In order to find an expression for $e^{\frac{i\hat{H}t}{\hbar}} $ we use $e^{x}  = \sum_{n=0}^{\infty} \frac{x^n}{n!}$ as follows
\begin{equation}
	\label{eqn:UDerive} 
	\begin{split}
	e^{\frac{i\hat{H}t}{\hbar}}  &= \sum_{n=0}^{\infty} \frac{\left( \frac{i\hat{H}t}{\hbar}\right)^n}{n!}\\
	\sum_{k=1}^{N}e^{\frac{i\hat{H}t}{\hbar}}\ket{V_k}\bra{V_k}  &= \sum_{k=1}^{N}\sum_{n=1}^{\infty} \frac{\left( \frac{i\hat{H}t}{\hbar}\right)^n}{n!}\ket{V_k}\bra{V_k}\\
	&= \sum_{k=1}^{N}\sum_{n=1}^{\infty} \frac{\left( \frac{i\lambda_k t}{\hbar}\right)^n}{n!}\ket{V_k}\bra{V_k} \tab{} \textrm{as } \hat{H}^n\ket{V_k} = \lambda_k^n \ket{V_k}\\
	\hat{U} = e^{\frac{i\hat{H}t}{\hbar}} &= \sum_{k=1}^{N} e^{\frac{i \lambda_k t}{\hbar}} \ket{V_k}\bra{V_k}
	\end{split} 
\end{equation}
where $\lambda_k$ are eigenvalues and $V_k$ are eigenvectors of $\hat{H}$.

The unitary time evolution operator $\hat{U}$ for the three-level system takes the form
\begin{equation}
	\label{eqn:3LvlU} 
	\hat{U} = 
	\begin{bmatrix}
	a_{11} & a_{12} & a_{13}\\
	a_{12} & a_{22} & -a_{12}^{*}\\
	a_{13} & -a_{12}^{*} & a_{11}^{*}\\
	\end{bmatrix},
\end{equation}
where the matrix elements $a_{ij}$ are functions of parameters $\Delta, \Omega_R, \Omega_G$ and time which takes the form
{\begin{eqnarray}
	\label{eqn:3LvlU_Appex}
	\begin{split}
	a_{11} &= \frac{\left(\Delta ^2+\Omega _G^2\right) \cos \left(\Omega _G t\right)-2 i \Delta  \Omega _G \sin \left(\Omega _G t\right)+\Omega _R^2}{2 \Omega _G^2}\\
	a_{12} &= \frac{\Omega _R \left(\Delta  \left(\cos \left(\Omega _G t\right)-1\right)-i \Omega _G \sin \left(\Omega _G t\right)\right)}{\sqrt{2} \Omega _G^2}\\
 	a_{13} &= \frac{\Omega _R^2 \left( \cos \left(\Omega _G t\right)-1 \right)}{2 \Omega _G^2}\\
 	a_{22} & = \frac{\Delta ^2+\Omega _R^2 \cos \left(\Omega _G t\right)}{\Omega _G^2}\\
 	\end{split}
\end{eqnarray}}

In the case of initially populated top state $\begin{bmatrix} 1 & 0 & 0\end{bmatrix}^T$, the population at the end of the Rabi pulse is shown in Equation \ref{eqn:3LvlPop}. In a similar way to the two-level case we find the three-level state vector at any time of the atom-EM interaction $\ket{\Psi(t)} = \hat{U} \ket{\Psi(0)}$ and the population of each level is
\begin{equation}
	\label{eqn:3LvlPop} 
	P = 
	\begin{bmatrix}
	\lvert\Psi_{\ket{+1}}\rvert^2 \\
	\lvert\Psi_{\ket{0}}\rvert^2 \\
	\lvert\Psi_{\ket{-1}}\rvert^2 \\
	\end{bmatrix}
	=
	\begin{bmatrix}
	\frac{\Delta ^4+6 \Delta ^2 \Omega _G^2+4 \Omega _R^2 \left(\Delta ^2+\Omega _G^2\right) \cos \left(\Omega _G t\right)+\left(\Delta ^2-\Omega _G^2\right){}^2 \cos \left(2 \Omega _G t\right)+\Omega _G^4+2 \Omega _R^4}{8 \Omega _G^4}\\
	\frac{\Omega _R^2 \left(\Delta ^2 \left(\cos \left(\Omega _G t\right)-1\right){}^2+\Omega _G^2 \sin ^2\left(\Omega _G t\right)\right)}{2 \Omega _G^4}\\
	\frac{\Omega _R^4\left(1 - \cos \left(\Omega _G t\right)\right)^2}{4 \Omega _G^4}\\
	\end{bmatrix}
\end{equation}

\begin{figure}[h!]
	\centering
	\includegraphics[width=0.8\linewidth]{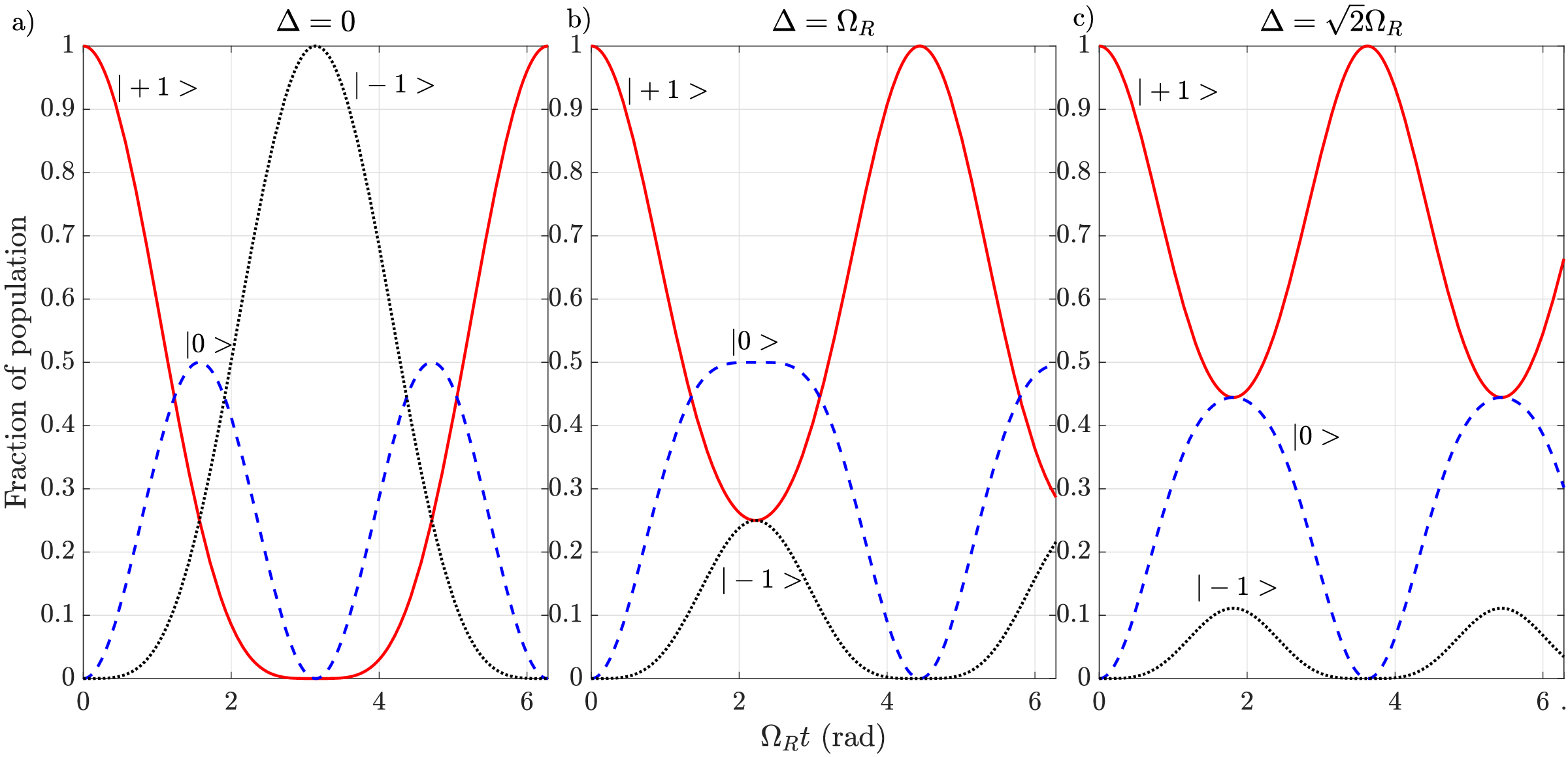}
	\caption[Rabi population for varying $\Omega_R$ and $\Delta$ based on analytical $\hat{U}$]{Population during the Rabi pulse in a three-level system for three combinations of $\Omega_R$ and $\Delta$ where the red solid, blue dashed and black dotted lines denote the populations of states $\ket{+1}, \ket{0}$ and $\ket{-1}$, respectively. Detuning are \textbf{a)} $\Delta = 0$, \textbf{b)} $\Delta = \Omega_R$ and \textbf{c)} $\Delta = \sqrt{2}\Omega_R$.}
	\label{fig:Analy_3Lvl}
\end{figure}

Figure \ref{fig:Analy_3Lvl} shows the population variation in a three-level system for three combinations of $\Omega_R$ and $\Delta$ during the Rabi pulse. The figure shows three interesting scenarios when the detuning $\Delta = 0$, $\Delta = \Omega_R$ and $\Delta = \sqrt{2}\Omega_R$. The key feature when $\Delta = \Omega_R$ is that $\lvert\Psi_{\ket{+1}}\rvert^2 = \lvert\Psi_{\ket{-1}}\rvert^2$ at $t=\frac{2.221}{\Omega_R}$. Further, when the detuning increases to $\Delta = \sqrt{2}\Omega_R$, $\lvert\Psi_{\ket{+1}}\rvert^2 = \lvert\Psi_{\ket{0}}\rvert^2$ at $t=\frac{1.814}{\Omega_R}$. Knowledge of these conditions is of great importance as it will assist in improving the stability of the splitting in a three-level Ramsey interferometer such as the $^{87}Rb$ $5^2S_{\frac{1}{2}} F = 1$ amidst experimental uncertainties in the applied EM field. 

For preliminary analysis the equal splitting of  $\lvert \Psi_{\ket{+1}}\rvert^2 = \lvert \Psi_{\ket{0}}\rvert^2$ at the $\Delta = 0$ condition is explored via the expressions in Equation \ref{eqn:3LvlPop} as the expected Ramsey signal for $\Delta = 0$ is non-oscillatory. When the equation $\lvert \Psi_{\ket{+1}}\rvert^2 - \lvert \Psi_{\ket{0}}\rvert^2 = 0$ for $\Delta = 0$ where $\Omega_G = \Omega_R$ is simplified the expression $\frac{1}{8} \left(4 \cos \left(\Omega _R t\right)+3 \cos \left(2 \Omega _R t\right)+1\right) = 0$ is achieved. When this is solved for $t$, $t = \frac{\arccos(\frac{1}{3}+2\pi c_1)}{\Omega _R}$, where $c_1$ is an integer and set at $c_1=0$ for the first equal splitting time where $t = \frac{\arccos(\frac{1}{3})}{\Omega _R}$. The general unitary time evolution operator $\hat{U}$ in Equation \ref{eqn:3LvlU} for the conditions $\Delta = 0$ and $t = \frac{\arccos(\frac{1}{3})}{\Omega _R}$ can be substituted leading to the much more simplified form of $\hat{U}_{\Delta=0}^{Split}$ shown below. Also, the free evolution operator $\hat{U}_{\Delta=0}^{Free}$ is achieved by substituting  $\Omega_R = \Omega_G = \Delta = 0$ to Equation \ref{eqn:3LvlU}.
\begin{equation}
	\label{eqn:3LvlU_D0} 
	\hat{U}_{\Delta=0}^{Split} = 
	\begin{bmatrix}
	\frac{2}{3} & -\frac{2 i}{3} & -\frac{1}{3} \\
	-\frac{2 i}{3} & \frac{1}{3} & -\frac{2 i}{3} \\
	-\frac{1}{3} & -\frac{2 i}{3} & \frac{2}{3} \\
	\end{bmatrix},
	\tab{}
	\hat{U}_{\Delta=0}^{Free} = 
	\begin{bmatrix}
	1 & 0 & 0 \\
	0 & 1 & 0 \\
	0 & 0 & 1\\
	\end{bmatrix}.
\end{equation}
By applying these unitary time evolution operations for each step of the Ramsey sequence, the wavefunction for the three-level system at the end of the sequence takes the form
\begin{equation}
	\label{eqn:3LvlUUUF} 
	\ket{\Psi_{sys}(t)} = \hat{U}^{Split}.\hat{U}^{Free}.\hat{U}^{Split}.\ket{\Psi(0)}.
\end{equation}
The resulting system wavefunction $\ket{\Psi_{sys}(t)}$ is shown below for the starting condition of $\ket{\Psi(0)} = F = \begin{bmatrix} 1 & 0 & 0\end{bmatrix}^T$. This can be easily converted to population which takes the form $P_{Rsy}$ as
\begin{equation}
	\label{eqn:3LvlU_Ramsey} 
	\ket{\Psi_{sys}(t)} = 
	\begin{bmatrix}
	\frac{1}{9}\\
	\frac{-4}{9} i \\
	\frac{-8}{9} \\
	\end{bmatrix},
	\tab{}
	P_{Rsy} = 
	\begin{bmatrix}
	\frac{1}{81}\\
	\frac{16}{81}\\
	\frac{64}{81}\\
	\end{bmatrix},
\end{equation}
where $	\ket{\Psi_{sys}(t)}$ and $P_{Rsy}$ are respectively the wavefunction and the population for the three-level system at the end of the Ramsey sequence. Here, the populations of states are at constant values of $P_{Rsy} = \begin{bmatrix} \frac{1}{81} & \frac{16}{81} & \frac{64}{81}\end{bmatrix} ^T$. The Ramsey signal can be obtained via the average spin projection for a multilevel system $\langle \hat{F}_Z \rangle = \hbar \sum_{m_F} m_F P_{m_F}$ where $P_{m_F}$ is the fractional population of the relevant $m_F$  state \cite{Anderson2010}. This leads to the constant value of $\frac{\langle \hat{F}_{Z}\rangle}{\hbar} = \frac{-7}{9}$ which is the expected behaviour of the system at $\Delta = 0$.

The scenario in Figure \ref{fig:Analy_3Lvl} \textbf{c)} is explored next as the Ramsey signal for $\Delta \neq 0$ is oscillatory and the equal splitting condition of  $\lvert \Psi_{\ket{+1}}\rvert^2 = \lvert \Psi_{\ket{0}}\rvert^2$ at $\Delta = \sqrt{2}\Omega_R$ is considered. When the condition $\lvert \Psi_{\ket{+1}}\rvert^2 - \lvert \Psi_{\ket{0}}\rvert^2 = 0$ for $\Delta = \sqrt{2}\Omega_R$ where $\Omega_G = \sqrt{3}\Omega_R$ is applied to Equation \ref{eqn:3LvlPop}, the expression $\frac{1}{6} \cos ^2\left(\frac{1}{2} \sqrt{3} \Omega _R t\right) \left(\cos \left(\sqrt{3} \Omega _R t\right)+5\right) = 0$ is achieved. When this is solved for $t$, $t = \frac{4 \pi  c_1+\pi }{\sqrt{3} \Omega _R}$ where $c_1$ is an integer and set at $c_1=0$ for the first equal splitting time where $t=\frac{\pi}{\sqrt{3} \Omega _R}$. The general unitary time evolution operator $\hat{U}$ in Equation \ref{eqn:3LvlU} for the conditions $\Delta = \sqrt{2}\Omega_R, \Omega_G = \sqrt{3}\Omega_R$ and $t = \frac{\pi}{\sqrt{3} \Omega _R}$ can be substituted and takes the much more simplified form of $\hat{U}_{\Delta=\sqrt{2}\Omega_R}^{Split}$ shown in Equation \ref{eqn:3LvlU_DR}. Further, the operator during free evolution is derived from Equations \ref{eqn:3LvlU} via Equations \ref{eqn:3LvlU_Appex} for the conditions $\Omega_R = 0$ when the general Rabi frequency $\Omega_G = \Delta$ for an evolution time $t=T$. This leads to the free evolution operator $\hat{U}_{\Omega_R=0}^{Free}$ in Equation \ref{eqn:3LvlU_DR},
\begin{equation}
	\label{eqn:3LvlU_DR} 
	\hat{U}_{\Delta=\sqrt{2}\Omega_R}^{Split} = 
	\begin{bmatrix}
	-\frac{2}{3} & -\frac{2}{3} & -\frac{1}{3} \\
	-\frac{2}{3} & \frac{1}{3} & \frac{2}{3} \\
	-\frac{1}{3} & \frac{2}{3} & -\frac{2}{3} \\
	\end{bmatrix}
	\tab{}
	\hat{U}_{\Omega_R=0}^{Free} = 
	\begin{bmatrix}
	e^{-i T \Delta } & 0 & 0 \\
	0 & 1 & 0 \\
	0 & 0 & e^{i T \Delta } \\
	\end{bmatrix}.
\end{equation}

Following Equation \ref{eqn:3LvlUUUF} for the full sequence, the wavefunction for the three-level system at the end of the Ramsey sequence takes the form for the condition of $\Delta = \sqrt{2}\Omega_R$
\begin{equation}
	\label{eqn:3LvlU_RamseyPsi} 
	\ket{\Psi_{sys}(T)} = 
	\begin{bmatrix}
	\frac{1}{9} (5 \cos (\Delta T )-3 i \sin (\Delta T )+4) \\
	\frac{2}{9} (\cos (\Delta T )-3 i \sin (\Delta T )-1) \\
	\frac{4}{9} (\cos (\Delta T )-1) \\
	\end{bmatrix},
\end{equation}

\begin{figure}[h!]
	\centering
	\includegraphics[width=0.8\linewidth]{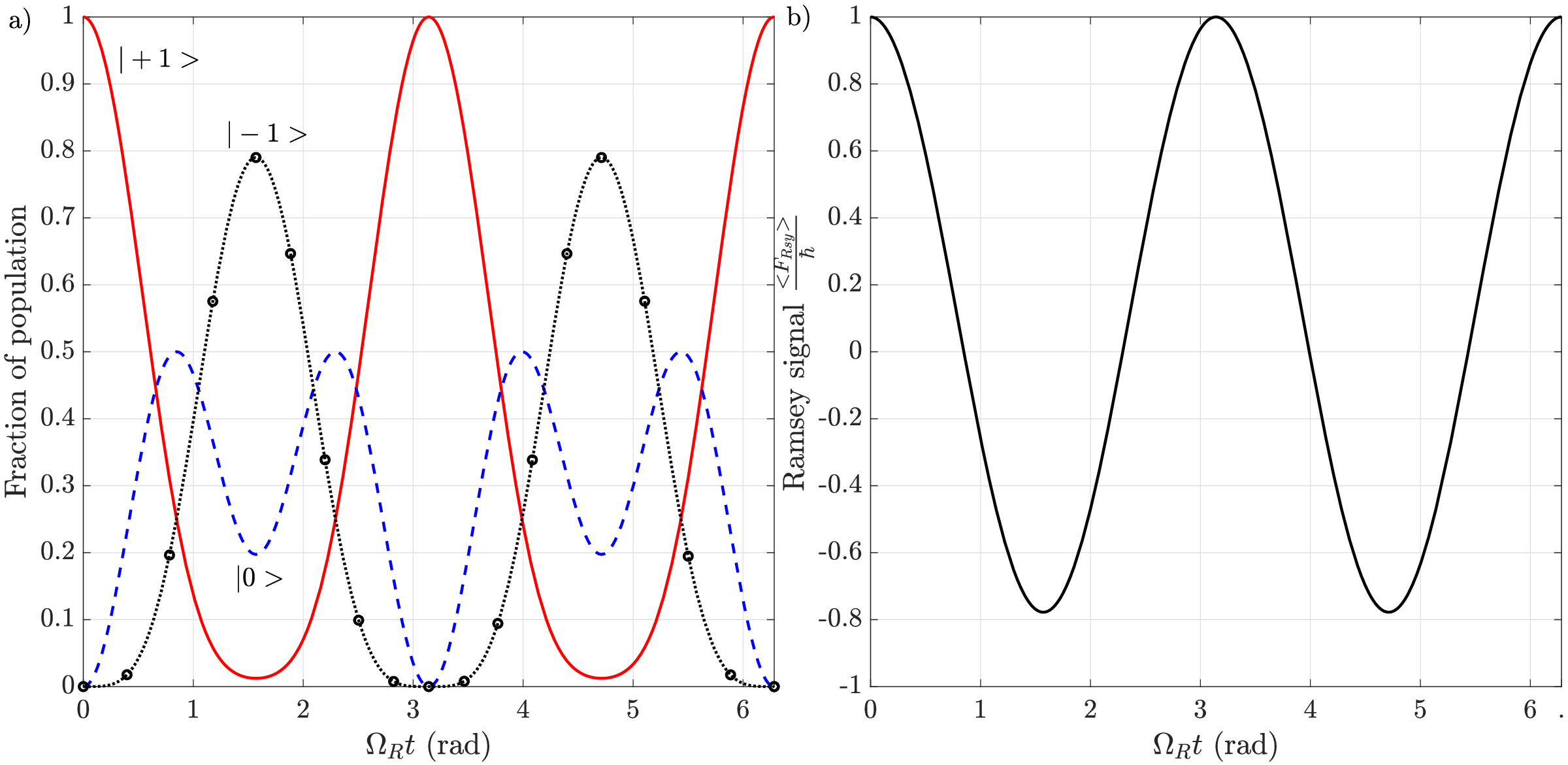}
	\caption[Analytical Ramsey expression of the three-level system for $\Delta = \sqrt{2}\Omega_R$]{Population variations and interference signal $\frac{\langle \hat{F}_{Z} \rangle}{\hbar}$ after the Ramsey sequence for the case $\Delta = \sqrt{2}\Omega_R$ in the three-level system. \textbf{a)} Variation of the population of each state where the red solid line denotes $\ket{+1}$, the blue dashed line denotes $\ket{0}$ and the black dotted line denote $\ket{-1}$. \textbf{b)} Variation of the Ramsey signal based on $\frac{\langle \hat{F}_Z \rangle}{\hbar} = \sum_{m_F} m_F P_{m_F}$, where $P_{m_F}$ is the fraction of population of state $\ket{m_F}$.}
	\label{fig:3Lvl_AnaDRRamsey}
\end{figure}

Further, an overall expression for the Ramsey signal can be obtained when the average spin projection for a multilevel system $\langle \hat{F}_Z \rangle = \hbar \sum_{m_F} m_F P_{m_F}$ is applied. Based on $\langle \hat{F}_Z \rangle$ the Ramsey signal takes the form $\frac{\langle \hat{F}_{Z} \rangle}{\hbar} = \frac{1}{9} (1 + 8 \cos (\Delta  T))$ where the Ramsey signal and populations of each state are shown in Figure \ref{fig:3Lvl_AnaDRRamsey}.

\pagebreak
\section{Rabi and Ramsey analytical models for five-level system}\label{five-level system}
\begin{figure}[h!]
	\centering
	\includegraphics[width=0.8\linewidth]{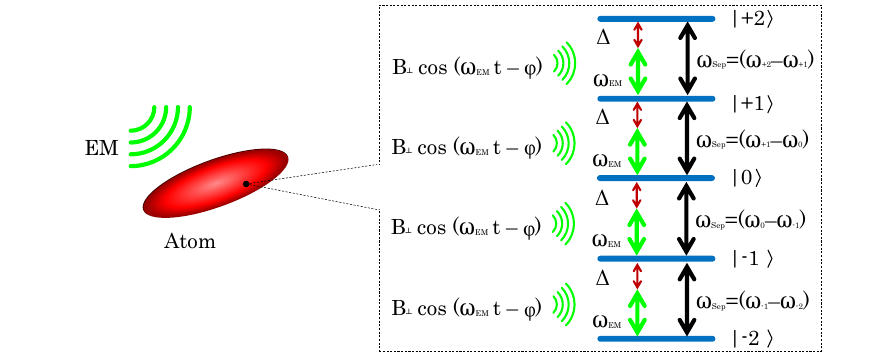}
	\caption[Schematic of five-level atom in an oscillating electromagnetic field]{Five-level atom interacting with an oscillating EM field where the energy separation between adjacent states is equal to $\hbar \omega_{Sep}$.}
	\label{fig:FiveLevelAtom}
\end{figure}

A schematic of the five-level system is shown in Figure \ref{fig:FiveLevelAtom} with equal energy separation between adjacent states leading to $\omega_{\ket{i}}-\omega_{\ket{i\pm1}}= \omega_{Sep}$. Consider this system being coupled to an external EM field with a frequency $\omega_{EM} = {\omega_{Sep}}-\Delta$ where $\Delta$ is the detuning. The Rabi frequency for this system is $\Omega_R = \frac{\mu_0 g_F B_{\perp}}{\hbar}$ \cite{Anderson2010} which is the magnetic dipole coupling between adjacent states.

\subsection{Rabi oscillations in the five-level system}\label{Rabi5Lvl}
The methodology presented for the three-level system is followed here to derive analytical expressions for the five-level Rabi and Ramsey signals. However, as a precursor for the five-level Rabi signal, \cite{Anderson2010} via \cite{Majorana1932, Vitanov1997}, shows that the analytical solutions for the transition probabilities of a spin-$F$ system with $2F+1$ sub-levels can be obtained via expressions for $C_1(t)$ and $C_2(t)$ in Equation \ref{eqn:Psi_Sys} for the two-level system as below;
\begin{equation}
	\label{eqn:Flvl}
	\begin{split}
	\Psi_{m_F} &= \sqrt{\frac{(2F)!}{(F+m_F)!(F-m_F)!}} C_1(t)^{F-m_F} C_2(t)^{F+m_F} \tab m_F \in \{-F \rightarrow F\}\\
	\lvert \Psi_{\ket{-2}} \rvert ^2 &= (|C_1|^2)^4, \\
	\lvert \Psi_{\ket{-1}} \rvert ^2 &= 4(|C_1|^2)^3(1-|C_1|^2), \\
	\lvert \Psi_{\ket{0}} \rvert ^2 &= 6(|C_1|^2)^2(1-|C_1|^2)^2, \\
	\lvert \Psi_{\ket{+1}} \rvert ^2 &= 4(|C_1|^2)(1-|C_1|^2)^3, \\
	\lvert \Psi_{\ket{+2}} \rvert ^2 &= (1-|C_1|^2)^4, 
	\end{split}
\end{equation}
where $C_1(t), C_2(t)$ are defined in Equation \ref{eqn:Psi_Sys} and $|C_1|^2 = |C_1(t)|^2 = \frac{\Omega_R^2}{\Omega^2_G} \sin^2(\frac{\Omega_G t}{2})$. 

\begin{figure}[h!]
	\centering
	\includegraphics[width=0.8\linewidth]{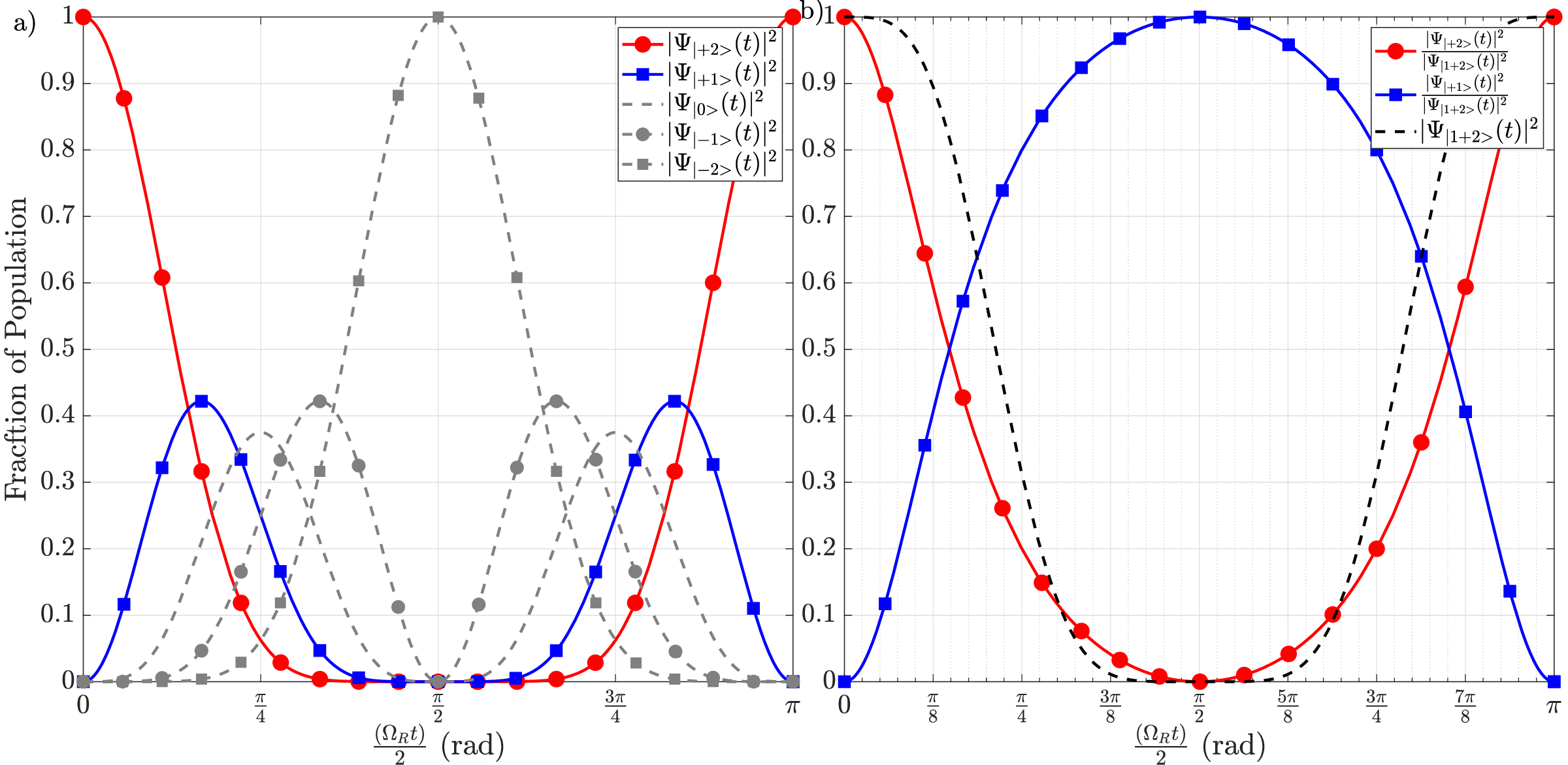}
	\caption[Analytical resonant Rabi oscillations within the $^{87}Rb 5^2S_{\frac{1}{2}} F = 2$]{Resonant Rabi oscillations in the $F = 2$ system according to Equation \ref{eqn:Flvl}. \textbf{a)} Evolution of populations of the five states with time, \textbf{b)} population fraction in state $\ket{+1}$ (blue solid line with squares) and in state $\ket{+2}$ (red solid line with circles) relative to the combined population of the two states $\ket{+1}$ and $\ket{+2}$. The black dashed line shows the combined population of states $\ket{+1}-\ket{+2}$.}
	\label{fig:FLvl}
\end{figure}

Based on this approach, the evolution of the fractional population of each level when the BEC starts in $\ket{+2}$ can be derived as shown in Equation \ref{eqn:Flvl}. It should be noted that this treatment can be applied to arbitrary values of $F$ and is only valid for linearly shifted Zeeman levels coupled via magnetic dipole transitions. Figure \ref{fig:FLvl} \textbf{a)} presents the on-resonance ($\Delta = 0$) analytical solutions for Equation \ref{eqn:Flvl} \cite{Anderson2010} showing the fractional population of each level with respect to the Rabi pulse area. The red line denotes the population variation of the state $\ket{+2}$ and the blue line of the state $\ket{+1}$. The analytical solution shows that the two levels create an equal population fraction of \num{0.41} at $\sin \left( \frac{\Omega_R t}{2} \right) = \frac{1}{\sqrt{5}}$, where $\frac{\Omega_R t}{2} = \SI{0.46}{\radian}$ or $\SI{2.68}{\radian}$. At this time \SI{18}{\percent} of the atoms from the total cloud are in the $\ket{0}$ and $\ket{-1}$ states. This is reflected in Figure \ref{fig:FLvl} \textbf{b)} which represents the fraction of atom numbers in state $\ket{+1}$ and $\ket{+2}$ with respect to the combined $\ket{+1}-\ket{+2}$ system. The red solid line with circles is the fractional population of $\ket{+2}$ and the blue solid line with squares is the fractional population of $\ket{+1}$. The black dashed line shows the dynamics of the combined $\ket{+1}-\ket{+2}$ system where the atoms move to other states of the five-level system. Figure \ref{fig:FLvl} \textbf{b)} indicates that any population mix where $\ket{+2} \in \{\SI{100}{\percent} \rightarrow \SI{50}{\percent}\}$ and $\ket{+1} \in \{\SI{0}{\percent} \rightarrow \SI{50}{\percent}\}$ is possible with minimal loss of atoms in the range $\frac{\Omega_R t}{2} \in \{\SI{0}{\radian} \rightarrow \SI{0.46}{\radian} \}$. However, going beyond this point the superposition leads to a fast decay of the signal where the atoms quickly move out of the desired $\ket{+1}-\ket{+2}$ states which recovers back when $\frac{\Omega_R t}{2} = \SI{2.68}{\radian}$. Between $\frac{\Omega_R t}{2} \in \{\SI{2.68}{\radian} \rightarrow \SI{3.14}{\radian}\}$ the other half of the combination of superpositions from $\ket{+2} \in \{\SI{50}{\percent} \rightarrow \SI{100}{\percent}\}$ and $\ket{+1} \in \{\SI{50}{\percent} \rightarrow \SI{0}{\percent}\}$ are available.

We will use the interaction Hamiltonian $\hat{H_I}$ for a five-level system in the form of Equation \ref{eqn:5LvlHI} which takes into account the Clebsch-Gordon coefficients for the corresponding $F = 2$ hyperfine manifold \cite{Anderson2010}. 
\begin{equation}
	\label{eqn:5LvlHI} 
	\hat{H}_I = \hbar 
	\begin{bmatrix}
	2 \Delta & \Omega _R & 0 & 0 & 0 \\
	\Omega _R & \Delta & \sqrt{\frac{3}{2}} \Omega _R & 0 & 0 \\
	0 & \sqrt{\frac{3}{2}} \Omega _R & 0 & \sqrt{\frac{3}{2}} \Omega _R & 0 \\
	0 & 0 & \sqrt{\frac{3}{2}} \Omega _R & -\Delta & \Omega _R \\
	0 & 0 & 0 & \Omega _R & -2 \Delta \\
	\end{bmatrix}
\end{equation}
where $\hbar$ is the reduced Planck's constant, $\Delta$ is the detuning of the external laser field from the energy separation between adjacent states and $\Omega_R = \frac{\mu_0 g_F B_{\perp}}{\hbar}$ \cite{Anderson2010} is the resonant Rabi frequency.

Again, using the standard method of solving the eigenvalue-eigenvector problem we solve for the normalised eigenvalues $\lambda_i$ and the eigenvectors $V_i$. Once the eigenvalues and eigenvectors are obtained, the unitary time evolution operator $\hat{U}$ for the general case is found via $\hat{U} = \sum_{i=1}^{n} e^{\frac{-i \lambda_i t}{\hbar}} \ket{V_i} \bra{V_i}$. The resulting bare matrix is sizeable beyond several pages. To simplify, the relation of $\Delta=\alpha\Omega_R$ (leading to $\alpha = \frac{\Delta}{\Omega_R}$) is introduced where $\Omega_G=\sqrt{1+\alpha^2}\Omega_R$. With this the evolution operator $\hat{U}$ takes the reduced form of
\begin{equation}
	\label{eqn:5LvlU} 
	\hat{U} = 
	\begin{bmatrix}
	a_{11} & a_{12} & a_{13} & a_{14} & a_{15}\\
	a_{12} & a_{22} & a_{23} & a_{24} & -a_{14}^*\\
	a_{13} & a_{23} & a_{33} & -a_{23}^* & a_{13}^*\\
	a_{14} & a_{24} & -a_{23}^* & a_{22}^* & -a_{12}^*\\
	a_{15} & -a_{14}^* & a_{13}^* & -a_{12}^* & a_{11}^*\\
	\end{bmatrix},
\end{equation}
where $a_{ij}^*$ is the complex conjugate and $a_{ij} = f(\alpha, \Omega_R, t)$ which take the following form;

\begin{equation}
	\begin{array}{l}
	\fl \tab{} a_{11} = \frac{\left(8 \alpha ^2+4\right) \cos \left(\Omega _G t\right)-8 i \alpha  \sqrt{\alpha ^2+1} \sin \left(\Omega _G t\right) \left(\left(2 \alpha ^2+1\right) \cos \left(\Omega _G t\right)+1\right)+\left(8 \left(\alpha ^4+\alpha ^2\right)+1\right) \cos \left(2 \Omega _G t\right)+3}{8 \left(\alpha ^2+1\right)^2}\\
	\fl \tab{} a_{12} = \frac{\left(4 \alpha ^2+3\right) \alpha  \cos \left(2 \Omega _G t\right)+i \left(-2 \sqrt{\alpha ^2+1} \sin \left(\Omega _G t\right) \left(-2 \alpha ^2+\left(4 \alpha ^2+1\right) \cos \left(\Omega _G t\right)+1\right)+3 i \alpha \right)-4 \alpha ^3 \cos \left(\Omega _G t\right)}{4 \left(\alpha ^2+1\right)^2}\\
	\fl \tab{} a_{13} = -\frac{\sqrt{\frac{3}{2}} \sin ^2\left(\frac{1}{2} \Omega _G t\right) \left(-2 i \alpha  \sqrt{\alpha ^2+1} \sin \left(\Omega _G t\right)+\left(2 \alpha ^2+1\right) \cos \left(\Omega _G t\right)+1\right)}{\left(\alpha ^2+1\right)^2}\\
 	\fl \tab{} a_{14} = \frac{2 \sin ^3\left(\frac{1}{2} \Omega _G t\right) \left(\alpha  \sin \left(\frac{1}{2} \Omega _G t\right)+i \sqrt{\alpha ^2+1} \cos \left(\frac{1}{2} \Omega _G t\right)\right)}{\left(\alpha ^2+1\right)^2}\\
 	\fl \tab{} a_{15} = \frac{\sin ^4\left(\frac{1}{2} \Omega _G t\right)}{\left(\alpha ^2+1\right)^2}\\
	\fl \tab{} a_{22} = \frac{\left(\alpha ^2+2 \cos \left(\Omega _G t\right)-1\right) \left(-2 i \alpha  \sqrt{\alpha ^2+1} \sin \left(\Omega _G t\right)+\left(2 \alpha ^2+1\right) \cos \left(\Omega _G t\right)+1\right)}{2 \left(\alpha ^2+1\right)^2}\\
 	\fl \tab{} a_{23} = -\frac{\sqrt{6} \sin \left(\frac{1}{2} \Omega _G t\right) \left(\alpha ^2+\cos \left(\Omega _G t\right)\right) \left(\alpha  \sin \left(\frac{1}{2} \Omega _G t\right)+i \sqrt{\alpha ^2+1} \cos \left(\frac{1}{2} \Omega _G t\right)\right)}{\left(\alpha ^2+1\right)^2}\\
 	\fl \tab{} a_{24} = -\frac{\sin ^2\left(\frac{1}{2} \Omega _G t\right) \left(3 \alpha ^2+2 \cos \left(\Omega _G t\right)+1\right)}{\left(\alpha ^2+1\right)^2}\\
 	\fl \tab{} a_{33} = \frac{\left(1-2 \alpha ^2\right)^2+12 \alpha ^2 \cos \left(\Omega _G t\right)+3 \cos \left(2 \Omega _G t\right)}{4 \left(\alpha ^2+1\right)^2}\\
 	\end{array} 
\end{equation}

The population at the end of the Rabi pulse can be obtained when this $\hat{U}$ is applied to the starting state of $\ket{+2}$ $F = \begin{bmatrix} 0 & 0 & 0 & 0 & 1 \end{bmatrix}^T$ where the population at the end of the pulse is 
\begin{equation}
	\label{eqn:5LvlPop_Gen} 
	P = 
	\begin{bmatrix}
	\lvert\Psi_{\ket{-2}}\rvert^2 \\
	\lvert\Psi_{\ket{-1}}\rvert^2 \\
	\lvert\Psi_{\ket{0}}\rvert^2 \\
	\lvert\Psi_{\ket{+1}}\rvert^2 \\
	\lvert\Psi_{\ket{+2}}\rvert^2 \\
	\end{bmatrix}
	=
	\begin{bmatrix}
	\frac{\sin ^8\left(\frac{1}{2} \Omega _G t\right)}{\left(\alpha ^2+1\right)^4} \\
	\frac{2 \left(2 \alpha ^2+\cos \left(\Omega _G t\right)+1\right) \sin ^6\left(\frac{1}{2} \Omega _G t\right)}{\left(\alpha ^2+1\right)^4} \\
	\frac{3 \left(2 \alpha ^2+\cos \left(\Omega _G t\right)+1\right){}^2 \sin ^4\left(\frac{1}{2} \Omega _G t\right)}{2 \left(\alpha ^2+1\right)^4} \\
	\frac{\left(2 \alpha ^2+\cos \left(\Omega _G t\right)+1\right){}^3 \sin ^2\left(\frac{1}{2} \Omega _G t\right)}{2 \left(\alpha ^2+1\right)^4} \\
	\frac{\left(2 \alpha ^2+\cos \left(\Omega _G t\right)+1\right){}^4}{16 \left(\alpha ^2+1\right)^4} \\
	\end{bmatrix}
\end{equation}
where $\Delta=\alpha\Omega_R$ and $\Omega_G=\sqrt{1+\alpha^2}\Omega_R$.

When the equal splitting condition is applied where $\lvert\Psi_{\ket{+2}}\rvert^2 = \lvert\Psi_{\ket{+1}}\rvert^2$ and solved for $t$, the equal splitting occurs at $t = \frac{4 \left(\tan ^{-1}\left(\sqrt{\frac{-\alpha ^2+2 \sqrt{5} \sqrt{4-\alpha ^2}+9}{\alpha ^2+1}}\right)+\pi  c_1\right)}{\Omega _G}$ where $c_1$ is an integer. The first equal splitting occurs at $c_1=0$ where $t_{Split} = \frac{4 \tan ^{-1}\left(\sqrt{\frac{-\alpha ^2+2 \sqrt{5} \sqrt{4-\alpha ^2}+9}{\alpha ^2+1}}\right)}{\Omega _G}$. When this condition is applied to $\hat{U}$ in Equation \ref{eqn:5LvlU}, the splitting matrix takes the form
\begin{equation}
\resizebox{0.9 \textwidth}{!} 
{$
	\label{eqn:5LvlU_Gen_EQSp} 
	\hat{U}_{G}^{Split} = 
	\begin{bmatrix}
	A_{11} & A_{12} & A_{13} & A_{14} & A_{15}\\
	A_{12} & A_{22} & A_{23} & A_{24} & -A_{14}^*\\
	A_{13} & A_{23} & A_{33} & -A_{23}^* & A_{13}^*\\
	A_{14} & A_{24} & -A_{23}^* & A_{22}^* & -A_{12}^*\\
	A_{15} & -A_{14}^* & A_{13}^* & -A_{12}^* & A_{11}^*\\
	\end{bmatrix}, 
	\hat{U}_{\Omega_R=0 }^{Free} = 
	\begin{bmatrix}
	e^{-2i t \Delta } & 0& 0 & 0 & 0\\
	0 & e^{-i t \Delta } & 0 & 0 & 0\\
	0 &	0 & 1 & 0 & 0\\
	0 & 0 & 0 & e^{i t \Delta } &0\\
	0 & 0 & 0 & 0 &e^{2i t \Delta }\\
	\end{bmatrix}
$}
\end{equation}
where $\hat{U}_G^{Split}$ is achieved by substituting  $t = t_{Split}$ to Equation \ref{eqn:5LvlU} where the full expressions are where $A_{ij} = f(\alpha)$ and the full expressions are as follows;

\begin{equation}
	\label{eqn:5LvlUGenSplit_Appex}
	\begin{array}{l}
	\fl \tab{}\tab{} A_{11} = \frac{4 \left(2 \left(\alpha ^8-2 \alpha ^6-5 \alpha ^4+2\right)+5 i \alpha  \left(\alpha ^2-2\right) \left(\alpha ^2+1\right)^{3/2} \sin \left(\Omega_G t_{Split} \right)\right)}{25 \left(\alpha ^2+1\right)^2}\\
	\fl \tab{}\tab{} A_{12} = \frac{8}{25}\frac{\alpha \left( \alpha^2+1 \right) \left( \alpha^2-3 \right) + 25 i \left(-\frac{\left(\alpha ^2-1\right) \left(\alpha ^2+1\right)^{5/2} \left(\alpha ^2-2 \sqrt{5} \sqrt{4-\alpha ^2}-9\right) \left(\alpha ^2-\sqrt{5} \sqrt{4-\alpha ^2}-4\right) \sqrt{\frac{-\alpha ^2+2 \sqrt{5} \sqrt{4-\alpha ^2}+9}{\alpha ^2+1}}}{\left(\sqrt{5} \sqrt{4-\alpha ^2}+5\right)^4} \right)}{\left(\alpha ^2+1\right)^2}\\
	\fl \tab{}\tab{} A_{13} = \frac{1}{25} \sqrt{6} \left(2 \alpha ^2+\frac{5 i \alpha  \sin \left(\Omega_G t_{Split}\right)}{\sqrt{\alpha ^2+1}}-4\right)\\
	\fl \tab{}\tab{} A_{14} = \frac{2 \sin ^3\left(\frac{\Omega_G t_{Split}}{2} 2\right) \left(\alpha  \sin \left(\frac{\Omega_G t_{Split}}{2}\right)+i \sqrt{\alpha ^2+1} \cos \left(\frac{\Omega_G t_{Split}}{2}\right)\right)}{\left(\alpha ^2+1\right)^2}\\
	\fl \tab{}\tab{} A_{15} = \frac{1}{25}\\
	\fl \tab{}\tab{} A_{22} = \frac{1}{25} \left(-2 \alpha ^2-\frac{5 i \alpha  \sin \left( \Omega_G t_{Split}\right)}{\sqrt{\alpha ^2+1}}+4\right)\\
	\fl \tab{}\tab{} A_{23} = \frac{3}{25} \sqrt{\frac{3}{2}} \left(-2 \alpha -\frac{5 i \sin \left(\Omega_G t_{Split}\right)}{\sqrt{\alpha ^2+1}}\right)\\
	\fl \tab{}\tab{} A_{24} = -\frac{11}{25}\\
	\fl \tab{}\tab{} A_{33} = \frac{1}{25}\\
 	\end{array} 
\end{equation}

One can easily obtain the expression for the system wavefunction at the end of the Ramsey sequence by applying $\hat{U}_{G}^{Split}, \hat{U}_{\Omega_R=0 }^{Free}$ to Equation \ref{eqn:3LvlUUUF} which then can be converted to populations and Ramsey signal via $\langle \hat{F}_Z \rangle = \hbar \sum_{m_F} m_F P_{m_F}$. However, these expressions are omitted due to the immense length.  These expressions reduce to a much more convenient form when a specific relation between $\Delta$ and $\Omega_R$ via $\Delta=\alpha\Omega_R$ is applied. Preliminarily, two conditions of $\Delta = 0$ and $\Delta = 2\Omega_R$ will be considered. In the first case, when $\Delta = 0 \rightarrow \Omega_G = \Omega_R$, which is substituted into $\hat{U}$ in Equation \ref{eqn:5LvlU} to obtain 
\begin{equation}
	\label{eqn:5LvlU_D0} 
	\hat{U}_{\Delta=0} = 
	\begin{bmatrix}
	b_{11} & b_{12} & b_{13} & b_{14} & b_{15}\\
	b_{12} & b_{22} & b_{23} & b_{24} & -b_{14}^*\\
	b_{13} & b_{23} & b_{33} & -b_{23}^* & b_{13}^*\\
	b_{14} & b_{24} & -b_{23}^* & b_{22}^* & -b_{12}^*\\
	b_{15} & -b_{14}^* & b_{13}^* & -b_{12}^* & b_{11}^*\\
	\end{bmatrix}
\end{equation}
where $b_{ij} = f(\Omega_R, t)$ and the full expressions are
\begin{equation}
	\label{eqn:5LvlUD0_Appex}
	\begin{array}{ll}
	b_{11} = \cos ^4\left(\frac{\Omega _R t}{2}\right) & 
	b_{22} = \frac{1}{2} \left(\cos \left(\Omega _R t\right)+\cos \left(2 \Omega _R t\right)\right)\\
	b_{12} = -\frac{1}{4} i \left(2 \sin \left(\Omega _R t\right)+\sin \left(2 \Omega _R t\right)\right) & 
	b_{23} = -\frac{1}{2} i \sqrt{\frac{3}{2}} \sin \left(2 \Omega _R t\right)\\
 	b_{13} = -\frac{1}{2} \sqrt{\frac{3}{2}} \sin ^2\left(\Omega _R t\right) &
 	b_{24} = \frac{1}{2} \left(\cos \left(2 \Omega _R t\right)-\cos \left(\Omega _R t\right)\right)\\
 	b_{14} = i \sin ^2\left(\frac{\Omega _R t}{2}\right) \sin \left(\Omega _R t\right) & 
 	b_{33} = \frac{1}{4} \left(3 \cos \left(2 \Omega _R t\right)+1\right)\\
 	b_{15} = \sin ^4\left(\frac{\Omega _R t}{2}\right)
 	\end{array}
\end{equation}

The population $P_{\Delta=0}$ at the end of the Rabi pulse for a specific starting state can be obtained by applying $\hat{U}_{\Delta=0}$ to the starting state. For the case of starting from the top state $\ket{+2}$ $F = \begin{bmatrix} 0 & 0 & 0 & 0 & 1 \end{bmatrix}^T$, the population at the end of the Rabi pulse is
\begin{equation}
	\label{eqn:5LvlPop_D0} 
	P_{\Delta=0} = 
	\begin{bmatrix}
	\lvert\Psi_{\ket{-2}}\rvert^2 \\
	\lvert\Psi_{\ket{-1}}\rvert^2 \\
	\lvert\Psi_{\ket{0}}\rvert^2 \\
	\lvert\Psi_{\ket{+1}}\rvert^2 \\
	\lvert\Psi_{\ket{+2}}\rvert^2 \\
	\end{bmatrix}
	=
	\begin{bmatrix}
	\sin ^8\left(\frac{\Omega _R t}{2}\right) \\
	\sin ^4\left(\frac{\Omega _R t}{2}\right) \sin ^2\left(\Omega _R t\right) \\
	\frac{3}{8} \sin ^4\left(\Omega _R t\right) \\
	\frac{1}{16} \left(2 \sin \left(\Omega _R t\right)+\sin \left(2 \Omega _R t\right)\right){}^2 \\
	\cos ^8\left(\frac{\Omega _R t}{2}\right)
	\end{bmatrix}
\end{equation}

\begin{figure}[h!]
	\centering
	\includegraphics[width=0.8\linewidth]{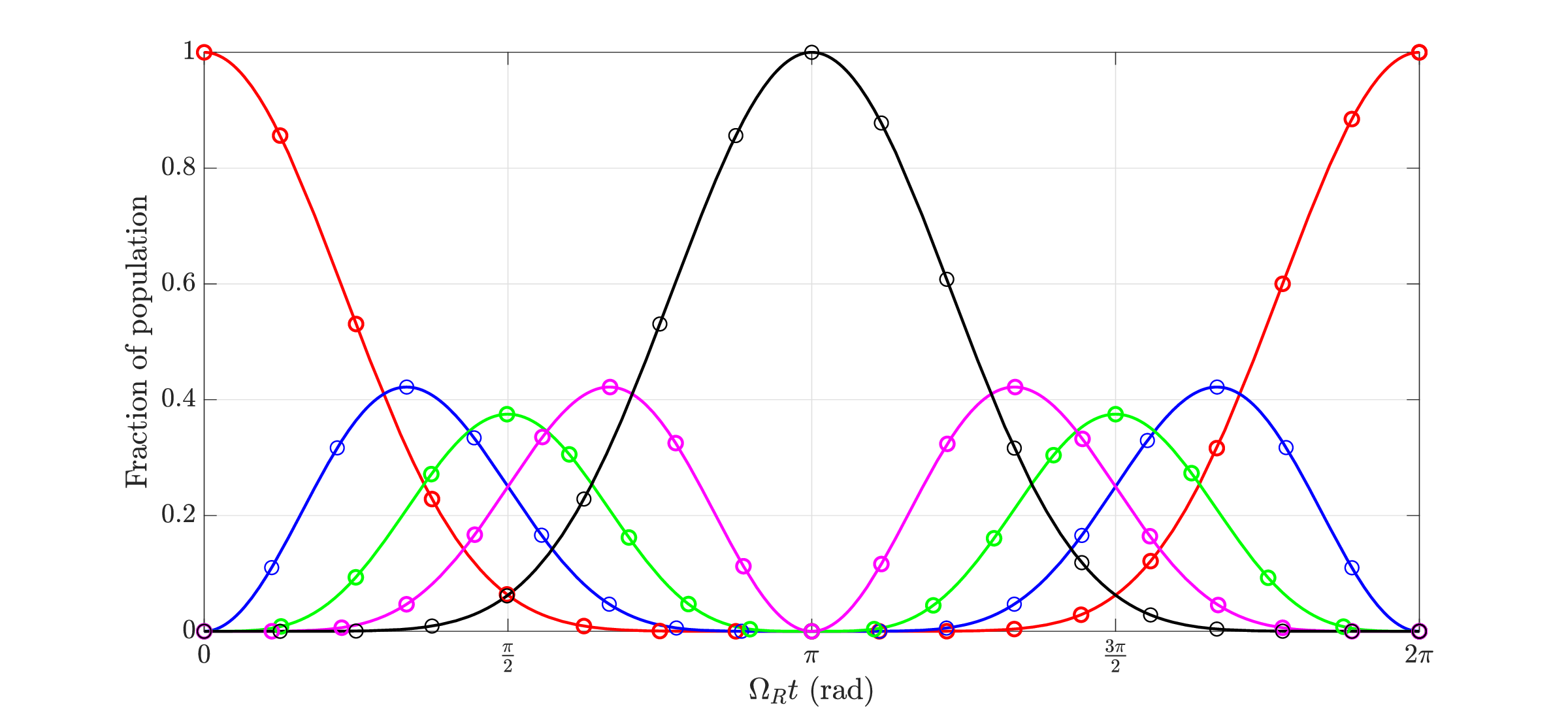}
	\caption[Analytical Rabi comparison of the five-level system for $\Delta = 0$]{Resonant ($\Delta=0$) Rabi oscillations of populations in the five-level system using the analytical form of Equation \ref{eqn:5LvlPop_D0} (solid lines) and of the $C_1, C_2$ representation of Equation \ref{eqn:Flvl} (open circles). Colours for states follow red - $\ket{+2}$, blue - $\ket{+1}$, green - $\ket{0}$, magenta - $\ket{-1}$ and black - $\ket{-2}$.}
	\label{fig:FLvl_AnaD0}
\end{figure}

These analytical formulae for the population after the Rabi pulse are compared with expressions in Equation \ref{eqn:Flvl} based on the two-level formalism as shown in Figure \ref{fig:FLvl_AnaD0}. The figure shows an exact overlap validating the expressions derived in Equation \ref{eqn:5LvlPop_D0}.

As shown above, during resonant Rabi interactions the atoms starting in state $\ket{+2}$ leave the state and occupy state $\ket{+1}$. As the EM pulse continues, this effect pushes atoms towards state $\ket{-2}$ via states $\ket{0}$ and $\ket{-1}$. Due to resonance, all atoms occupy $\ket{-2}$ and the system undergoes full inversion after which the atoms recover back to the initial state $\ket{+2}$ and continue to produce Rabi oscillations.

Moving on, the speciality of $\Delta = 2\Omega_R$ is that the two states of interest $\ket{+1}$ and $\ket{+2}$ do not cross when $\Delta$ increases beyond this point eliminating the possibility of creating a superposition with equal splitting. When $\Delta = 2\Omega_R \rightarrow \Omega_G = \sqrt{5}\Omega_R$ is substituted into $\hat{U}$ in Equation \ref{eqn:5LvlU} the $\hat{U}_{\Delta=2\Omega_R}$ we obtain
\begin{equation}
	\label{eqn:5LvlU_DR} 
	\hat{U} = 
	\begin{bmatrix}
	c_{11} & c_{12} & c_{13} & c_{14} & c_{15}\\
	c_{12} & c_{22} & c_{23} & c_{24} & -c_{14}^*\\
	c_{13} & c_{23} & c_{33} & -c_{23}^* & c_{13}^*\\
	c_{14} & c_{24} & -c_{23}^* & c_{22}^* & -c_{12}^*\\
	c_{15} & -c_{14}^* & c_{13}^* & -c_{12}^* & c_{11}^*\\
	\end{bmatrix},
\end{equation}
where $c_{ij} = f(\Omega_R, t)$ and the full expressions are
\begin{equation}
	\label{eqn:5LvlUDR_Appex}
	\begin{array}{l}
	\fl \tab{} \tab{} c_{11} = \frac{1}{200} \left(-16 i \sqrt{5} \sin \left(\sqrt{5} \Omega _R t\right)-72 i \sqrt{5} \sin \left(2 \sqrt{5} \Omega _R t\right)+36 \cos \left(\sqrt{5} \Omega _R t\right)+161 \cos \left(2 \sqrt{5} \Omega _R t\right)+3\right)\\
	\fl \tab{} \tab{} c_{12} = \frac{1}{100} \left(-17 i \sqrt{5} \sin \left(2 \sqrt{5} \Omega _R t\right)+14 i \sqrt{5} \sin \left(\sqrt{5} \Omega _R t\right)-32 \cos \left(\sqrt{5} \Omega _R t\right)+38 \cos \left(2 \sqrt{5} \Omega _R t\right)-6\right)\\
	\fl \tab{} \tab{} c_{13} = \frac{1}{25} \sqrt{\frac{3}{2}} \sin ^2\left(\frac{1}{2} \sqrt{5} \Omega _R t\right) \left(4 i \sqrt{5} \sin \left(\sqrt{5} \Omega _R t\right)-9 \cos \left(\sqrt{5} \Omega _R t\right)-1\right)\\
 	\fl \tab{} \tab{} c_{14} = \frac{2}{25} \sin ^3\left(\frac{1}{2} \sqrt{5} \Omega _R t\right) \left(2 \sin \left(\frac{1}{2} \sqrt{5} \Omega _R t\right)+i \sqrt{5} \cos \left(\frac{1}{2} \sqrt{5} \Omega _R t\right)\right)\\
 	\fl \tab{} \tab{} c_{15} = \frac{1}{25} \sin ^4\left(\frac{1}{2} \sqrt{5} \Omega _R t\right)\\
	\fl \tab{} \tab{} c_{22} = \frac{1}{50} \left(2 \cos \left(\sqrt{5} \Omega _R t\right)+3\right) \left(-4 i \sqrt{5} \sin \left(\sqrt{5} \Omega _R t\right)+9 \cos \left(\sqrt{5} \Omega _R t\right)+1\right)\\
 	\fl \tab{} \tab{} c_{23} = \frac{1}{50} \sqrt{\frac{3}{2}} \left(-8 i \sqrt{5} \sin \left(\sqrt{5} \Omega _R t\right)-i \sqrt{5} \sin \left(2 \sqrt{5} \Omega _R t\right)+12 \cos \left(\sqrt{5} \Omega _R t\right)+2 \cos \left(2 \sqrt{5} \Omega _R t\right)-14\right)\\
 	\fl \tab{} \tab{} c_{24} = \frac{1}{50} \left(11 \cos \left(\sqrt{5} \Omega _R t\right)+\cos \left(2 \sqrt{5} \Omega _R t\right)-12\right)\\
 	\fl \tab{} \tab{} c_{33} = \frac{1}{100} \left(48 \cos \left(\sqrt{5} \Omega _R t\right)+3 \cos \left(2 \sqrt{5} \Omega _R t\right)+49\right)\\
 	\end{array} 
\end{equation}

The population $P_{\Delta=2\Omega_R}$ at the end of the Rabi pulse for a specific starting state can be obtained by applying $\hat{U}_{\Delta=2\Omega_R}$ to the starting state. For the case of starting from the top state when $F = \begin{bmatrix} 0 & 0 & 0 & 0 & 1 \end{bmatrix}^T$, the population at the end of the Rabi pulse is
\begin{equation}
	\label{eqn:5LvlPop_DR} 
	P_{\Delta=2\Omega_R} = 
	\begin{bmatrix}
	\lvert\Psi_{\ket{-2}}\rvert^2 \\
	\lvert\Psi_{\ket{-1}}\rvert^2 \\
	\lvert\Psi_{\ket{0}}\rvert^2 \\
	\lvert\Psi_{\ket{+1}}\rvert^2 \\
	\lvert\Psi_{\ket{+2}}\rvert^2 \\
	\end{bmatrix}
	=
	\begin{bmatrix}
	\frac{1}{625} \sin ^8\left(\frac{1}{2} \sqrt{5} \Omega _R t\right) \\
	\frac{2}{625} \sin ^6\left(\frac{1}{2} \sqrt{5} \Omega _R t\right) \left(\cos \left(\sqrt{5} \Omega _R t\right)+9\right) \\
	\frac{3 \sin ^4\left(\frac{1}{2} \sqrt{5} \Omega _R t \right) \left(\cos \left(\sqrt{5} \Omega _R t\right)+9\right){}^2}{1250} \\
	\frac{\sin ^2\left(\frac{1}{2} \sqrt{5} \Omega _R t\right) \left(\cos \left(\sqrt{5} \Omega _R t\right)+9\right){}^3}{1250} \\
	\frac{\left(\cos \left(\sqrt{5} \Omega _R t\right)+9\right){}^4}{10000}
	\end{bmatrix}
\end{equation}

\begin{figure}[h!]
	\centering
	\includegraphics[width=0.8\linewidth]{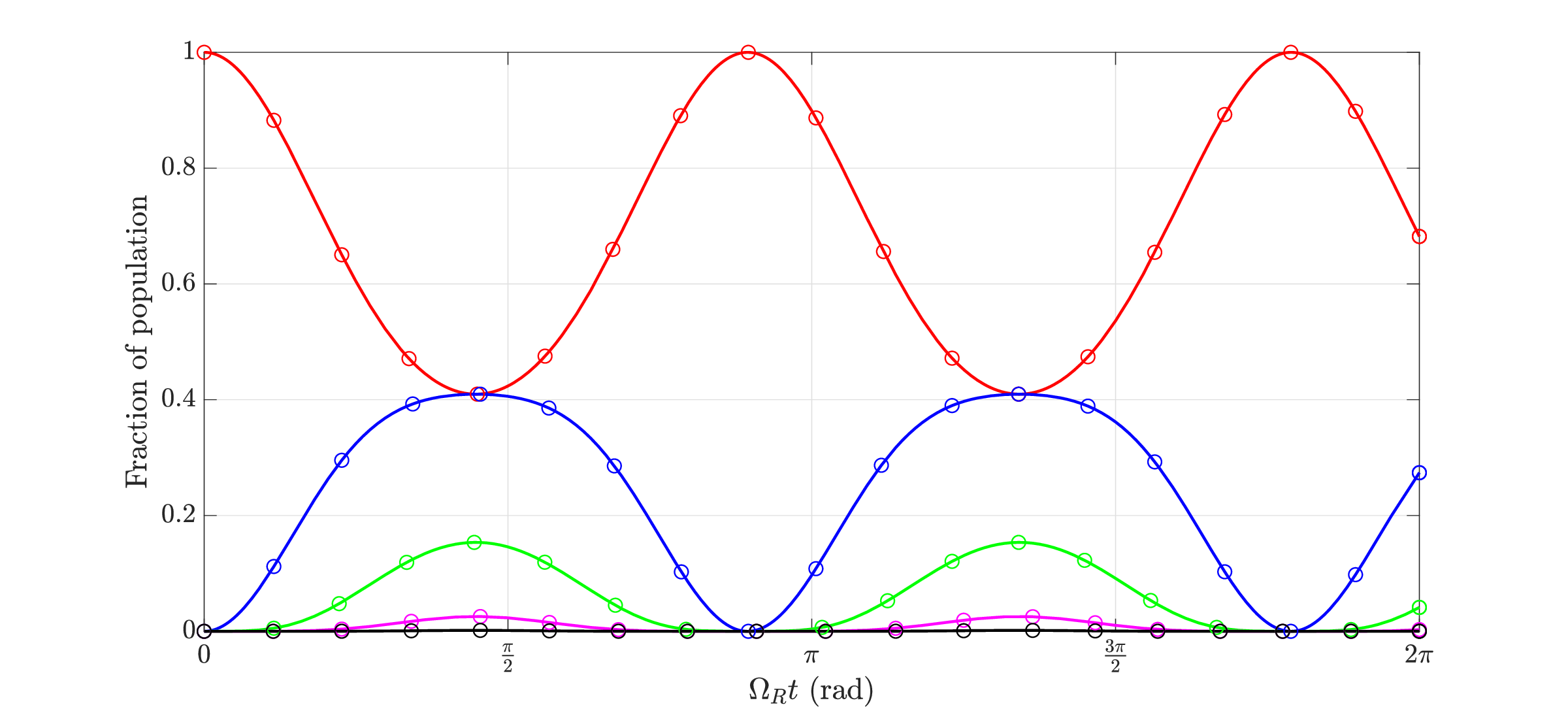}
	\caption[Analytical Rabi comparison of the five-level system for $\Delta = 2\Omega_R$.]{Non-resonant ($\Delta=2\Omega_R$) Rabi oscillations of populations in the five-level system using the analytical form of Equation \ref{eqn:5LvlPop_DR} (solid lines) and of the $C_1, C_2$ presentation of Equation \ref{eqn:Flvl} (open circles). Colours for states follow red - $\ket{+2}$, blue - $\ket{+1}$, green - $\ket{0}$, magenta - $\ket{-1}$ and black - $\ket{-2}$.}
	\label{fig:FLvl_AnaDR}
\end{figure}

Rabi oscillations of the populations show non-resonant behaviour when the initially populated state $\ket{+2}$ exhibits variations in the range \num{1} - \num{0.41} and state $\ket{+1}$- in the range \num{0} - \num{0.41} as shown in Figure \ref{fig:FLvl_AnaDR}. All other states show significantly smaller variations.

\subsection{Ramsey signal for the two interesting cases}\label{RamseyExpressions}
Firstly, for the resonant case where $\Delta = 0$, we are interested in the equal splitting of states $\ket{+2}$ and $\ket{+1}$. The equal population condition is applied between states $\ket{+2}$ and $\ket{+1}$ to obtain an expression for the pulse duration. When $\lvert \Psi_{\ket{+2}} \rvert^2 = \lvert \Psi_{\ket{+1}} \rvert^2$ is applied to Equation \ref{eqn:5LvlPop_D0}, $-\frac{1}{4} \sin ^2\left(\Omega _R t\right)-\frac{1}{16} \sin ^2\left(2 \Omega _R t\right)-\frac{1}{4} \sin \left(\Omega _R t\right) \sin \left(2 \Omega _R t\right)+\cos ^8\left(\frac{\Omega _R t}{2}\right) = 0$ is achieved where the acceptable terms for $t$ take the form $t=\frac{4 \left(\pi  c_1+\tan ^{-1}\left(2+\sqrt{5}\right)\right)}{\Omega _R}$ or $t=-\frac{4 \left(\pi  c_1+\tan ^{-1}\left(2-\sqrt{5}\right)\right)}{\Omega _R}$, where $c_1$ is an integer. The first equal splitting occurs at $c_1=0$ where $t= -\frac{4 \left(\tan ^{-1}\left(2-\sqrt{5}\right)\right)}{\Omega _R}$. When this condition is applied to $\hat{U}_{\Delta=0}$ in Equation \ref{eqn:5LvlU_D0} the desired time evolution operator $\hat{U}_{\Delta=0 }^{Split}$for equal splitting is obtained.
\begin{equation}
\resizebox{0.9 \textwidth}{!} 
{$
	\label{eqn:5LvlU_D0_EQSp} 
	\hat{U}_{\Delta=0 }^{Split} = 
	\begin{bmatrix}
	\frac{16}{25} & \frac{16 i}{25} & -\frac{4 \sqrt{6}}{25} & -\frac{4 i}{25} & \frac{1}{25} \\
	\frac{16 i}{25} & \frac{4}{25} & \frac{6 i \sqrt{6}}{25} & -\frac{11}{25} & -\frac{4 i}{25} \\
	-\frac{4 \sqrt{6}}{25} & \frac{6 i \sqrt{6}}{25} & \frac{1}{25} & \frac{6 i \sqrt{6}}{25} & -\frac{4 \sqrt{6}}{25} \\
 -\frac{4 i}{25} & -\frac{11}{25} & \frac{6 i \sqrt{6}}{25} & \frac{4}{25} & \frac{16 i}{25} \\
	\frac{1}{25} & -\frac{4 i}{25} & -\frac{4 \sqrt{6}}{25} & \frac{16 i}{25} & \frac{16}{25} \\
	\end{bmatrix}, 
	\hat{U}_{\Delta=0 }^{Free} = 
	\begin{bmatrix}
	1 & 0& 0 & 0 & 0\\
	0 & 1 & 0 & 0 & 0\\
	0 &	0 & 1 & 0 & 0\\
	0 & 0 & 0 & 1 & 0\\
	0 & 0 & 0 & 0 & 1\\
	\end{bmatrix}
$}
\end{equation}
where $\hat{U}_{\Delta=0 }^{Split}$ is achieved by substituting  $\Delta = 0, \Omega_G = \Omega_R$ and $t = -\frac{4 \left(\tan ^{-1}\left(2-\sqrt{5}\right)\right)}{\Omega _R}$ to Equation \ref{eqn:5LvlU} and $\hat{U}_{\Delta=0 }^{Free}$ is achieved by extrapolating from Equation \ref{eqn:3LvlU_D0}.

Following on from Equation \ref{eqn:3LvlU_D0}, the time evolution operator during free evolution $\hat{U}_{\Delta=0 }^{Free}$ is the identity matrix as shown in Equation \ref{eqn:5LvlU_D0_EQSp}. The equation for the system wavefunction at the end of the Ramsey sequence can be obtained by applying Equation \ref{eqn:3LvlUUUF} 
\begin{equation}
	\label{eqn:5LvlU_D0_Ramsey} 
	\ket{\Psi_{sys}(t)} = 
	\begin{bmatrix}
	\frac{81}{625} \\
	\frac{216 i}{625} \\
	-\frac{144 \sqrt{6}}{625} \\
	-\frac{384 i}{625} \\
	\frac{256}{625} \\
	\end{bmatrix},
	\tab{}
	P_{Rsy} = 
	\begin{bmatrix}
	\frac{65536}{390625} \\
	\frac{147456}{390625} \\
	\frac{124416}{390625} \\
	\frac{46656}{390625} \\
	\frac{6561}{390625} \\
	\end{bmatrix},
\end{equation}
where $	\ket{\Psi_{sys}(t)}$ and $P_{Rsy}$ are respectively the vector state and the populations for the five-level system at the end of the Ramsey sequence.

This can be easily converted to population which takes the form $P_{Rsy}$ in Equation \ref{eqn:5LvlU_D0_Ramsey} which shows constant values for the populations of each state at  $P_{Rsy} = \begin{bmatrix} \frac{65536}{390625} & \frac{147456}{390625} & \frac{124416}{390625} & \frac{46656}{390625} & \frac{6561}{390625} \end{bmatrix} ^T$. Further, an overall expression for the Ramsey signal can be obtained when the average spin projection for a multilevel system $\langle \hat{F}_Z \rangle = \hbar \sum_{m_F} m_F P_{m_F}$ is applied, where $P_{m_F}$ is the fractional population of the relevant state \cite{Anderson2010}. This leads to a constant value for the Ramsey signal at $\frac{\langle \hat{F}_Z \rangle}{\hbar} = \frac{-14}{25}$ which is the expected behaviour of the system.

For the second scenario, the detuning $\Delta = 2\Omega_R$ was chosen and when equal population condition of $\lvert \Psi_{\ket{+2}} \rvert^2 = \lvert \Psi_{\ket{+1}} \rvert^2$ is applied to the expressions Equation \ref{eqn:5LvlPop_DR}, leads to $\frac{\left(\cos \left(\sqrt{5} \Omega _R t\right)+9\right){}^4}{10000}-\frac{\sin ^2\left(\frac{1}{2} \sqrt{5} \Omega _R t\right) \left(\cos \left(\sqrt{5} \Omega _R t\right)+9\right){}^3}{1250} = 0$. Here, the acceptable term for $t$ takes the form $t=\frac{4 \pi  c_1+\pi }{\sqrt{5} \Omega _R}$, where $c_1$ is an integer. The first equal splitting occurs at $c_1=0$ where $t= \frac{\pi }{\sqrt{5} \Omega _R}$. When this condition is applied to $\hat{U}_{\Delta=2\Omega_R}$ in Equation \ref{eqn:5LvlU_DR} the desired time evolution operator $\hat{U}_{\Delta=2\Omega_R }^{Split}$ for equal splitting is obtained.
\begin{equation}
\resizebox{0.9 \textwidth}{!} 
{$
	\label{eqn:5LvlU_DR_EQSp} 
	\hat{U}_{\Delta=2\Omega_R }^{Split} = 
	\begin{bmatrix}
	\frac{16}{25} & \frac{16}{25} & \frac{4 \sqrt{6}}{25} & \frac{4}{25} & \frac{1}{25} \\
	\frac{16}{25} & -\frac{4}{25} & -\frac{6 \sqrt{6}}{25} & -\frac{11}{25} & -\frac{4}{25} \\
	\frac{4 \sqrt{6}}{25} & -\frac{6 \sqrt{6}}{25} & \frac{1}{25} & \frac{6 \sqrt{6}}{25} & \frac{4 \sqrt{6}}{25} \\
	\frac{4}{25} & -\frac{11}{25} & \frac{6 \sqrt{6}}{25} & -\frac{4}{25} & -\frac{16}{25} \\
	\frac{1}{25} & -\frac{4}{25} & \frac{4 \sqrt{6}}{25} & -\frac{16}{25} & \frac{16}{25} \\
	\end{bmatrix}, 
	\hat{U}_{\Omega_R=0 }^{Free} = 
	\begin{bmatrix}
	e^{-2i t \Delta } & 0& 0 & 0 & 0\\
	0 & e^{-i t \Delta } & 0 & 0 & 0\\
	0 &	0 & 1 & 0 & 0\\
	0 & 0 & 0 & e^{i t \Delta } &0\\
	0 & 0 & 0 & 0 &e^{2i t \Delta }\\
	\end{bmatrix},
$}
\end{equation}
where $\hat{U}_{\Delta=2\Omega_R }^{Split}$ is achieved by substituting  $\Delta = 2\Omega_R, \Omega_G = \sqrt{5}\Omega_R$ and $t =\frac{\pi }{\sqrt{5} \Omega _R}$ into Equation \ref{eqn:5LvlU}. Following on from Equation \ref{eqn:3LvlU_D0}, the time evolution operator during free evolution $\hat{U}_{\Omega_R=0 }^{Free}$ is extrapolated as shown in Equation \ref{eqn:5LvlU_D0_EQSp}.

The equation for the system wavefunction at the end of the Ramsey sequence can be obtained by applying Equation \ref{eqn:3LvlUUUF}
\begin{equation}
\resizebox{0.9 \textwidth}{!} 
{$
	\label{eqn:5LvlU_DR_Ramsey} 
	\ket{\Psi_{sys}(t)} = 
	\begin{bmatrix}
	\frac{256}{625} \sin ^4\left(\frac{\Delta  t}{2}\right)\\
	\frac{128}{625} \sin ^3\left(\frac{\Delta  t}{2}\right) \left(3 \sin \left(\frac{\Delta  t}{2}\right)+5 i \cos \left(\frac{\Delta  t}{2}\right)\right)\\
	\frac{4\sqrt{6}}{625} (5 i \sin (\Delta  t)-3 \cos (\Delta  t)+3)^2\\
	\frac{-4 e^{-2 i \Delta  t}}{625} \left(-1+e^{i \Delta  t}\right) \left(4+e^{i \Delta  t}\right)^3\\
	\frac{e^{-2 i \Delta  t}}{625} \left(4+e^{i \Delta  t}\right)^4
	\end{bmatrix},
	P_{Rsy} = 
	\begin{bmatrix}
	\frac{65536}{390625}\sin ^8\left(\frac{\Delta  t}{2}\right)\\
	\frac{16384}{390625}\sin ^6\left(\frac{\Delta  t}{2}\right) (8 \cos (\Delta  t)+17)\\
	\frac{1536}{390625}\sin ^4\left(\frac{\Delta  t}{2}\right) (8 \cos (\Delta  t)+17)^2\\
	\frac{64}{390625}\sin ^2\left(\frac{\Delta  t}{2}\right) (8 \cos (\Delta  t)+17)^3\\
	\frac{1}{390625}(8 \cos (\Delta  t)+17)^4
	\end{bmatrix}
$}
\end{equation}
where $\ket{\Psi_{sys}(t)}$ and $P_{Rsy}$ are respectively the vector state and the populations for the five-level system at the end of the Ramsey sequence.

\begin{figure}[h!]
	\centering
	\includegraphics[width=0.8\linewidth]{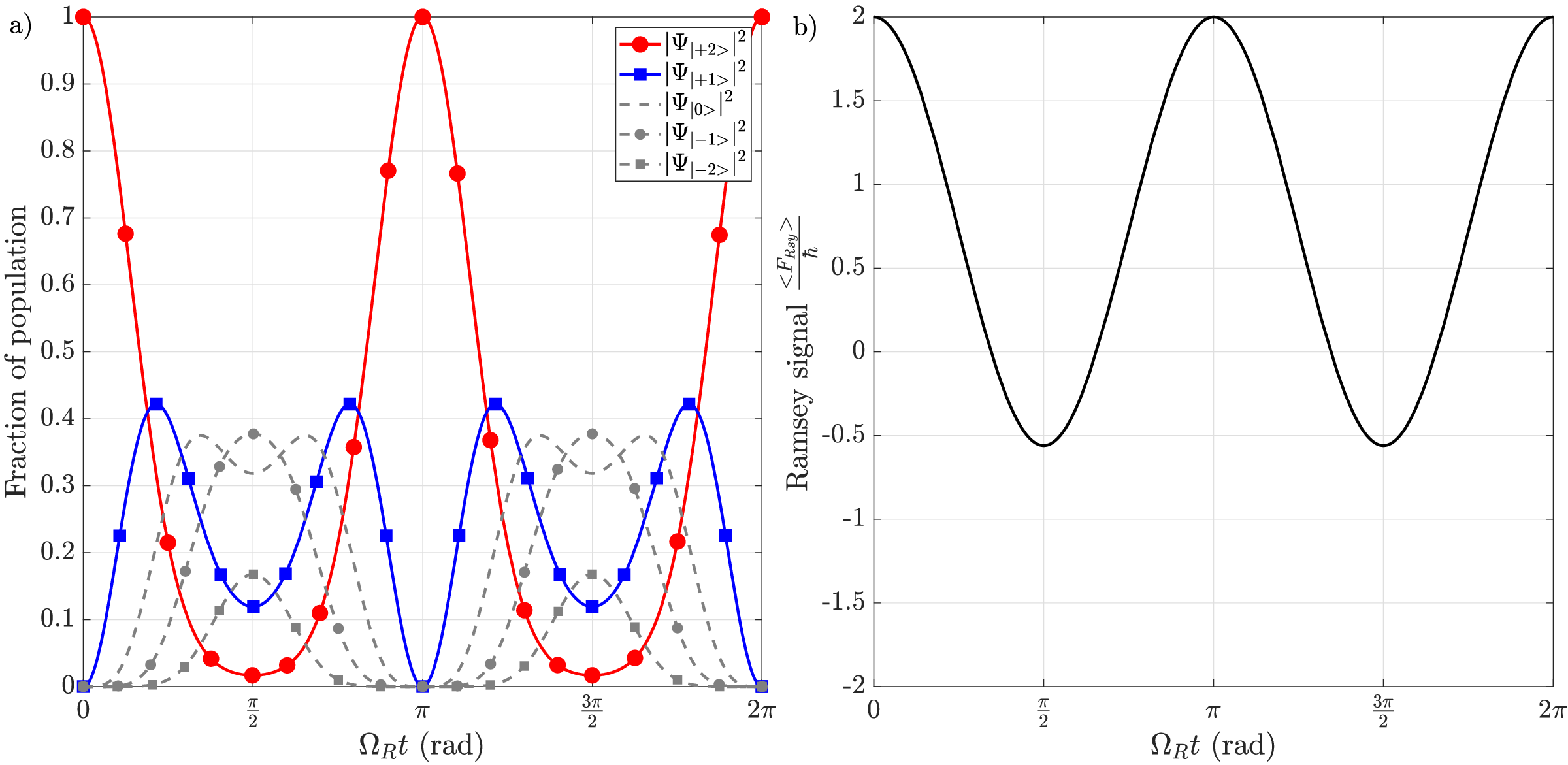}
	\caption[Analytical Ramsey interference of the five-level system for $\Delta = 2\Omega_R$]{ Population variations and the interference signal $\frac{\langle \hat{F}_Z \rangle}{\hbar}$ after the Ramsey sequence for the case of $\Delta = 2 \Omega_R$ in the five-level system. \textbf{a)}: The variation of the population of each state where the red solid line denotes $\ket{+2}$ and the blue solid line denotes $\ket{+1}$. \textbf{b)} The variation of the Ramsey signal based on $\frac{\langle \hat{F}_Z \rangle}{\hbar} = \sum_{m_F} m_F P_{m_F}$ where $P_{m_F}$ the is population fraction of state $\ket{m_F}$.}
	\label{fig:FLvl_AnaDRRamsey}
\end{figure}

This can be converted to populations in Equation \ref{eqn:5LvlU_DR_Ramsey}. Further, the Ramsey signal can be obtained when the average spin projection for a multilevel system $\langle \hat{F}_Z \rangle = \hbar \sum_{m_F} m_F P_{m_F}$ is applied \cite{Anderson2010}. Based on $\langle \hat{F}_Z \rangle$ the Ramsey signal takes the form $\frac{\langle \hat{F}_Z \rangle}{\hbar} = \frac{2}{25} (9 + 16 \cos (\Delta  t))$. The importance of $\Delta = 2\Omega_R$ scenario is that it denotes greater stability of the Rabi splitting after the pulse duration $t = \frac{\pi}{\sqrt{5}\Omega_R}$ under experimental phase and pulse uncertainties. This is visible in Figure \ref{fig:FLvl_AnaDR} showing minimal population variations of states $\ket{+1}$ and $\ket{+2}$ in the vicinity of equal splitting due to the flatness of the curves. Figure \ref{fig:FLvl_AnaDRRamsey} shows the population variation and the Ramsey signal for $\Delta = 2\Omega_R$. A decreased interference fringe contrast is noticeable compared to a near-resonant case but we anticipate the measured interference signal will be more stable in the presence of magnetic and frequency noise.

\subsection{Generalised Ramsey signal for the five-level system}\label{Ramsey5Lvl}
As shown under Equation \ref{eqn:5LvlU}, the general form of the unitary time evolution operator $\hat{U}$ can be derived which uses the important relation of $\Delta = \alpha \Omega_R$ and the population at the end of the first splitting pulse is shown in Equation \ref{eqn:5LvlPop_Gen}. Now, we look to expand the expression for $\hat{U}$ (Equation \ref{eqn:5LvlU}) by generalising the variations in the pulse duration. This is important as it allows to account for the experimental uncertainty in pulse which leads to variations in the population splitting at both the EM pulses of the Ramsey sequence. To do so, the pulse duration can be defined as $t=\frac{\beta}{\Omega_R}$, where $\beta$ is a phase relating to the pulse area defining the splitting of the system. This results in the following unitary time evolution operator
\begin{equation}
	\label{eqn:5LvlU_Gen_EQSp} 
	\hat{U}_{\alpha \beta}^{GSplit} = 
	\begin{bmatrix}
	B_{11} & B_{12} & B_{13} & B_{14} & B_{15}\\
	B_{12} & B_{22} & B_{23} & B_{24} & -B_{14}^*\\
	B_{13} & B_{23} & B_{33} & -B_{23}^* & B_{13}^*\\
	B_{14} & B_{24} & -B_{23}^* & B_{22}^* & -B_{12}^*\\
	B_{15} & -B_{14}^* & B_{13}^* & -B_{12}^* & B_{11}^*\\
	\end{bmatrix},
\end{equation}

where $\hat{U}_{\alpha \beta}^{GSplit}$ is achieved by substituting $t = \frac{\beta}{\Omega _R}$ into Equation \ref{eqn:5LvlU}, where the full expressions are
\begin{equation}
	\label{eqn:5LvlUAB_Appex}
	\begin{array}{l}
	\fl \tab{} \tab{} B_{11} = \frac{\left(8 \alpha ^2+4\right) \cos \left(\sqrt{\alpha ^2+1} \beta \right)-8 i \alpha  \sqrt{\alpha ^2+1} \sin \left(\sqrt{\alpha ^2+1} \beta \right) \left(\left(2 \alpha ^2+1\right) \cos \left(\sqrt{\alpha ^2+1} \beta \right)+1\right)+\left(8 \left(\alpha ^4+\alpha ^2\right)+1\right) \cos \left(2 \sqrt{\alpha ^2+1} \beta \right)+3}{8 \left(\alpha ^2+1\right)^2}\\
	\fl \tab{} \tab{} B_{12} = \frac{\left(4 \alpha ^2+3\right) \alpha  \cos \left(2 \sqrt{\alpha ^2+1} \beta \right)+i \left(-2 \sqrt{\alpha ^2+1} \sin \left(\sqrt{\alpha ^2+1} \beta \right) \left(\left(4 \alpha ^2+1\right) \cos \left(\sqrt{\alpha ^2+1} \beta \right)-2 \alpha ^2+1\right)+3 i \alpha \right)-4 \alpha ^3 \cos \left(\sqrt{\alpha ^2+1} \beta \right)}{4 \left(\alpha ^2+1\right)^2}\\
	\fl \tab{} \tab{} B_{13} = -\frac{\sqrt{\frac{3}{2}} \sin ^2\left(\frac{1}{2} \sqrt{\alpha ^2+1} \beta \right) \left(-2 i \alpha  \sqrt{\alpha ^2+1} \sin \left(\sqrt{\alpha ^2+1} \beta \right)+\left(2 \alpha ^2+1\right) \cos \left(\sqrt{\alpha ^2+1} \beta \right)+1\right)}{\left(\alpha ^2+1\right)^2}\\
 	\fl \tab{} \tab{} B_{14} = \frac{2 \sin ^3\left(\frac{1}{2} \sqrt{\alpha ^2+1} \beta \right) \left(\alpha  \sin \left(\frac{1}{2} \sqrt{\alpha ^2+1} \beta \right)+i \sqrt{\alpha ^2+1} \cos \left(\frac{1}{2} \sqrt{\alpha ^2+1} \beta \right)\right)}{\left(\alpha ^2+1\right)^2}\\
 	\fl \tab{} \tab{} B_{15} = \frac{\sin ^4\left(\frac{1}{2} \sqrt{\alpha ^2+1} \beta \right)}{\left(\alpha ^2+1\right)^2}\\
 	\fl \tab{} \tab{} B_{22} = \frac{\left(2 \cos \left(\sqrt{\alpha ^2+1} \beta \right)+\alpha ^2-1\right) \left(-2 i \alpha  \sqrt{\alpha ^2+1} \sin \left(\sqrt{\alpha ^2+1} \beta \right)+\left(2 \alpha ^2+1\right) \cos \left(\sqrt{\alpha ^2+1} \beta \right)+1\right)}{2 \left(\alpha ^2+1\right)^2}\\
 	\fl \tab{} \tab{} B_{23} = -\frac{\sqrt{6} \sin \left(\frac{1}{2} \sqrt{\alpha ^2+1} \beta \right) \left(\cos \left(\sqrt{\alpha ^2+1} \beta \right)+\alpha ^2\right) \left(\alpha  \sin \left(\frac{1}{2} \sqrt{\alpha ^2+1} \beta \right)+i \sqrt{\alpha ^2+1} \cos \left(\frac{1}{2} \sqrt{\alpha ^2+1} \beta \right)\right)}{\left(\alpha ^2+1\right)^2}\\
 	\fl \tab{} \tab{} B_{24} = -\frac{\sin ^2\left(\frac{1}{2} \sqrt{\alpha ^2+1} \beta \right) \left(2 \cos \left(\sqrt{\alpha ^2+1} \beta \right)+3 \alpha ^2+1\right)}{\left(\alpha ^2+1\right)^2}\\
 	\fl \tab{} \tab{} B_{33} = \frac{12 \alpha ^2 \cos \left(\sqrt{\alpha ^2+1} \beta \right)+3 \cos \left(2 \sqrt{\alpha ^2+1} \beta \right)+\left(1-2 \alpha ^2\right)^2}{4 \left(\alpha ^2+1\right)^2}\\
 	\end{array} 
\end{equation}

The wavefunction at the end of the Ramsey sequence can be obtained when $\hat{U}_{\alpha \beta}^{GSplit}$ is applied to Equation \ref{eqn:3LvlUUUF}, which can be converted to the population, which are omitted due to the immense lengths of the expressions. However, an expression for the Ramsey signal can be obtained when these populations are subjected to $\frac{\langle \hat{F}_Z \rangle}{\hbar} = \sum_{m_F} m_F P_{m_F}$, which results in a complete analytical expression for the average spin at the end of Ramsey interference in the five-level model:
\begin{eqnarray}
	\label{eqn:5Lvl_Ramsey}
	\fl
\frac{\langle \hat{F}_Z \rangle}{\hbar} &= \frac{4 \sqrt{\alpha ^2+1} \alpha ^2 \cos \left(\sqrt{\alpha ^2+1} \beta \right)+\sqrt{\alpha ^2+1} \cos \left(2 \sqrt{\alpha ^2+1} \beta \right)+\left(2 \alpha ^4+1\right) \sqrt{\alpha ^2+1}}{\left(\alpha ^2+1\right)^{5/2}} 
\nonumber \\ \fl
& - \frac{4 \sin ^2\left(\frac{1}{2} \sqrt{\alpha ^2+1} \beta \right)\left(\sqrt{\alpha ^2+1} \left(\left(2 \alpha ^2+1\right) \cos \left(\sqrt{\alpha ^2+1} \beta \right)+1\right)\right)}{\left(\alpha ^2+1\right)^{5/2}} \cos (\text{$\Delta $t}) \nonumber \\ \fl
& + \frac{4 \sin ^2\left(\frac{1}{2} \sqrt{\alpha ^2+1} \beta \right)\left(2 \left(\alpha ^3+\alpha \right)\sin \left(\sqrt{\alpha ^2+1} \beta \right)\right)}{\left(\alpha ^2+1\right)^{5/2}}  \sin (\text{$\Delta $t}),
\end{eqnarray}
where $\alpha = \frac{\Delta}{\Omega_R}$ and $\beta = \Omega_R t$.

By using the trigonometric conversion of $a \cos(\theta) + b \sin(\theta) = A \cos(\theta-\phi)$, Equation \ref{eqn:5Lvl_Ramsey} can be further simplified to the rather elegant form of: 

\begin{eqnarray}
	\label{eqn:5Lvl_Ramsey_Elegant}
\fl
\frac{\langle \hat{F}_Z \rangle}{\hbar} &= \frac{4 \sqrt{\alpha ^2+1} \alpha ^2 \cos \left(\sqrt{\alpha ^2+1} \beta \right)+\sqrt{\alpha ^2+1} \cos \left(2 \sqrt{\alpha ^2+1} \beta \right)+\left(2 \alpha ^4+1\right) \sqrt{\alpha ^2+1}}{\left(\alpha ^2+1\right)^{5/2}} 
\nonumber\\ \fl
 &- \frac{4 \sin ^2\left(\frac{1}{2} \sqrt{\alpha ^2+1} \beta \right) \left(\sqrt{\left(\alpha ^2+1\right)} \left(\cos \left(\sqrt{\alpha ^2+1} \beta \right)+2 \alpha ^2+1\right)\right)}{\left(\alpha ^2+1\right)^{5/2}} \cos (\text{$\Delta $t}-\phi ).
\end{eqnarray}
where $\tan(\phi) = \frac{-2 \left(\alpha ^3+\alpha \right) \sin \left(\sqrt{\alpha ^2+1} \beta \right)}{\sqrt{\alpha ^2+1} \left(\left(2 \alpha ^2+1\right) \cos \left(\sqrt{\alpha ^2+1} \beta \right)+1\right)}$.

The prominent feature is that Equation \ref{eqn:5Lvl_Ramsey_Elegant} reduces to $\frac{\langle \hat{F}_Z \rangle}{\hbar} = A-B \cos(\Delta t - \phi)$  when values for $\alpha$ and $\beta$ are substituted. Further, the unique relation of $A + B = 2$ is also reported. As a special note, $\beta$ scans only within the first half of one Rabi cycle in the above analytical description. However, we scan the Rabi signal beyond one cycle in experiments; therefore, $t = \frac{(2\pi + \beta)}{\Omega_R}$ or $t = \frac{(4\pi - \beta)}{\Omega_R}$ should be used when converting a fitted $\beta$ to experimental results.

\section{Discussion and conclusions}
Here we have explored the analytical description of three- and five-level systems via unitary time evolution operators and obtained analytical expressions for describing the Ramsey interferometric signal for a typical Ramsey sequence. Several interesting Rabi oscillations for the three-level system are shown in Figure \ref{fig:Analy_3Lvl}. Further, the behaviour of Rabi oscillations of the five-level system is verified via the expansion of expressions from the two-level model as shown in Figures \ref{fig:FLvl_AnaD0} and \ref{fig:FLvl_AnaDR}. Several special cases and examples of how these analytical expressions can be used to obtain population variations along with the averaged Ramsey signal at the end of the Ramsey sequence are also presented as shown in Figures \ref{fig:3Lvl_AnaDRRamsey} and \ref{fig:FLvl_AnaDRRamsey}. Finally, a generalised equation for the average Ramsey signal at the standard Ramsey sequence for the five-level system is presented in Equation \ref{eqn:5Lvl_Ramsey_Elegant} where the splitting condition is also generalised expanding the applicability.

A limitation of this analysis is that both splitting pulses of the Ramsey sequence are considered to be equal. However, this analysis can be expanded to Ramsey sequences with unequal splitting pulses by following the derivation methodology presented here. This means that a separate unitary time evolution operator $\hat{U}$ is to be derived for the second pulse. From this the analytical expression for the system wavefunction at the end of the Ramsey sequence can be obtained via Equation \ref{eqn:3LvlUUUF}. From this the populations of each state can be obtained from which the analytical expression for Ramsey interference can be obtained via $\langle \hat{F}_Z \rangle = \hbar \sum_{m_F} m_F P_{m_F}$ where $P_{m_F}$ is the fractional population of the relevant $m_F$  state \cite{Anderson2010}. Similarly, this same methodology can be used to obtain analytical expressions for Ramsey sequences with spin-echo techniques such as for work reported in \cite{Eto2013, Egorov2011, Opanchuk2012}. For this, a new unitary time evolution operator $\hat{U}^{\pi}$ is to be derived for the $\pi-$ pulse. Once Equation \ref{eqn:3LvlUUUF} is expanded to $\ket{\Psi_{sys}(t)} = \hat{U}^{Split}.\hat{U}^{Free}.\hat{U}^{\pi}.\hat{U}^{Free}.\hat{U}^{Split}.\ket{\Psi(0)}$; the analytical expression for the system wavefunction is obtained. From here, the analytical expression for the Ramsey interference is obtained by converting the system wavefunction to the multilevel populations.

All in all, a comprehensive analysis of the three- and five-level systems via the unitary time evolution operator under the equal-Rabi condition for a Ramsey sequence with equal splitting pulses is discussed.

\section*{References}
\bibliography{References}

\end{document}